\documentclass[superscriptaddress,floatfix,amsmath,notitlepage,twocolumn]{revtex4-1}
\usepackage{graphicx}
\usepackage{color}
\usepackage{siunitx}
\usepackage[normal]{subfigure}
\usepackage[usenames,dvipsnames]{xcolor}
\usepackage{amssymb}
\usepackage{dsfont}
\newcommand{\be}{\begin{equation}}
\newcommand{\ee}{\end{equation}}

\usepackage[colorlinks=true,urlcolor=blue,linkcolor=blue,citecolor=blue]{hyperref}

\begin{document}

\title{Dynamics of many-body photon bound states in chiral waveguide QED}

\author{Sahand Mahmoodian}
\affiliation{Institute for Theoretical Physics, Institute for Gravitational Physics (Albert Einstein Institute), Leibniz University Hannover, Appelstra{\ss}e 2, 30167 Hannover, Germany}
\author{Giuseppe Calaj\'o}
\affiliation{ICFO-Institut de Ciencies Fotoniques, The Barcelona Institute of
Science and Technology, 08860 Castelldefels (Barcelona), Spain}
\author{Darrick E. Chang}
\affiliation{ICFO-Institut de Ciencies Fotoniques, The Barcelona Institute of
Science and Technology, 08860 Castelldefels (Barcelona), Spain}
\affiliation{ICREA-Instituci\'{o}  Catalana  de  Recerca  i  Estudis  Avan\c{c}ats,  08015  Barcelona,  Spain}
\author{Klemens Hammerer}
\affiliation{Institute for Theoretical Physics, Institute for Gravitational Physics (Albert Einstein Institute), Leibniz University Hannover, Appelstra{\ss}e 2, 30167 Hannover, Germany}
\author{Anders~S.~S{\o}rensen}
\affiliation{Center for Hybrid Quantum Networks (Hy-Q), Niels Bohr Institute, University of Copenhagen, Blegdamsvej 17, DK-2100 Copenhagen, Denmark}

\date{\today}
 
\begin{abstract}
We theoretically study the few- and many-body dynamics of photons in chiral waveguides. In particular, we examine pulse propagation through an ensemble of $N$ two-level systems chirally coupled to a waveguide. We show that the system supports correlated multi-photon bound states, which have a well-defined photon number $n$ and propagate through the system with a group delay scaling as $1/n^2$. This has the interesting consequence that, during propagation, an incident coherent-state pulse breaks up into different bound-state components that can become spatially separated at the output in a sufficiently long system. For sufficiently many photons and sufficiently short systems, we show that linear combinations of $n$-body bound states recover the well-known phenomenon of mean-field solitons in self-induced transparency. Our work thus covers the entire spectrum from few-photon quantum propagation, to genuine quantum many-body (atom and photon) phenomena, and ultimately the quantum-to-classical transition.  Finally, we demonstrate that the bound states can undergo elastic scattering with additional photons. Together, our results demonstrate that photon bound states are truly distinct physical objects emerging from the most elementary light-matter interaction between photons and two-level emitters. Our work opens the door to studying quantum many-body physics and soliton physics with photons in chiral waveguide QED.
\end{abstract}

\maketitle

\section{Introduction}

Generating quantum many-body states of light remains one of the outstanding challenges of modern quantum optics \cite{Chang2014NPHOT}. Such many-body states of light are of fundamental physical interest as they arise from non-equilibrium systems with strong interactions between light and matter. On the other hand, they also promise to form novel resources for quantum technologies, for example, in quantum-enhanced metrology \cite{Paulisch2019PRA}. The main obstacle in the pursuit of generating such many-body states has been the difficulty in developing one-dimensional systems with a sufficiently strong nonlinear response at the few-photon scale \cite{Chang2018RMP}.  Recently however, significant progress has been made in creating an ideal light--matter interface between atoms or artificial emitters coupled to a one-dimensional continuum of photons at optical \cite{Arcari2014PRL, Goban2014NCOM, Lodahl2015RMP} and microwave frequencies \cite{Bishop2009NPHYS, Hoi2012PRL, Mirhosseini2019Nature}. Such an interface creates a highly nonlinear medium as photons propagating in a waveguide interact deterministically with atoms. Systems of this kind have thus far been used to propose or demonstrate the generation of states of photons with strong two- or three-body correlations \cite{Birnbaum2005Nature, Faraon2008NPHYS, Reinhard2012NPHOT, Peyronel2012Nature, Firstenberg2013Nature, Javadi2015NCOM, Hamsen2017PRL, DeSantis2017NNANO,  Liang2017Science, Mahmoodian2018PRL, Stiesdal2018PRL}. 

Studies of photon correlations in these systems typically consider steady-state driving and photon correlations are subsequently measured in relative coordinates. On the other hand, here we show that pulse propagation through two-level systems (TLSs) strongly coupled to a waveguide leads to very distinct temporal features which reveal the underlying dynamics and has the potential to generate temporally ordered many-body states of light. In particular we theoretically consider the propagation of pulses of coherent and Fock states of light through a waveguide to which $N$ TLSs are chirally coupled. We show that photons undergo a Wigner delay \cite{Wigner1955PR, BourgainOptLett2013} --- a delay of the pulse center due the optical excitation transferring to the atom and back to the waveguide --- which is dependent on the number of photons.
\begin{figure}[!t]
\includegraphics[width=\columnwidth]{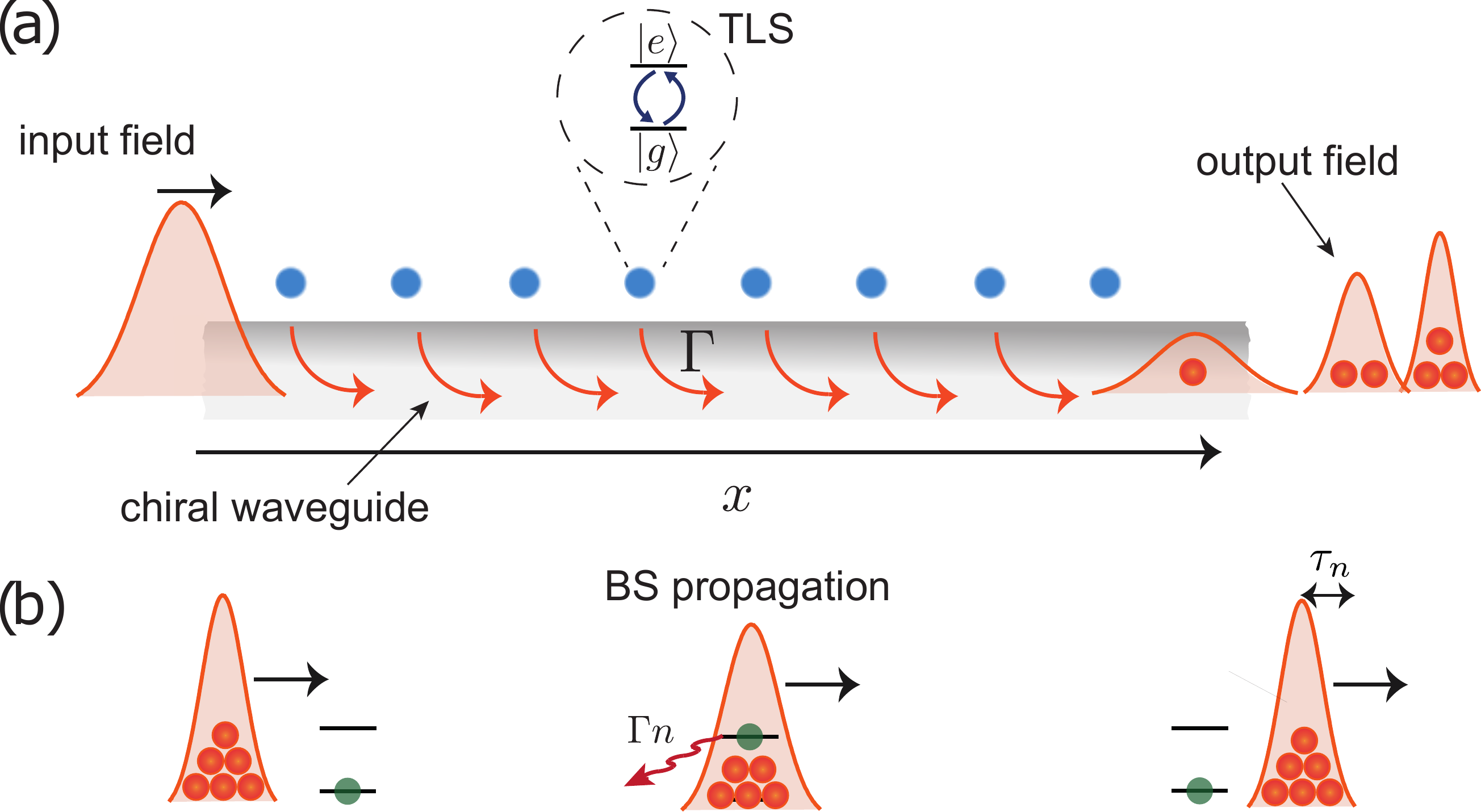}
\caption{\label{fig:schem}(a)  $N$ two-level atoms (blue circles) are chirally coupled to a waveguide with decay rate $\Gamma$ and driven by an input Gaussian pulse, which can be a coherent or Fock state. The light pulse propagates with a correlation-number-dependent group velocity leading to an output state where one-, two- and three-photon bound states are spatially separated. (b) Schematic of the bound states propagation. When an $n$-photon bound state is scattered by an atom, it re-emits the absorbed photon with a stimulated emission rate  $\Gamma n$, coinciding with the inverse of its width.  Since only a single photon out of $n$ is delayed by an amount $4/(\Gamma n)$, the pulse preserves its shape but is delayed by $\tau_n = 4/( \Gamma n^2)$.}
\end{figure}
Therefore, incident pulses break up into a state that is temporally ordered by its photon correlations. For example, as shown in Fig.~\ref{fig:schem}(a), an incident coherent field can produce a pulse with three-photon correlations followed by two-photon correlations states followed by uncorrelated photons. The underlying physics that causes this time delay is examined by considering the many-body photonic scattering eigenstates of TLSs chirally coupled to a waveguide.

Of central importance are the class of bound eigenstates, where two or more photons propagate together. We show that the photon bound states propagate past each atom with a photon-number-dependent delay of $\tau_n = 4/(n^2\Gamma)$, where $\Gamma$ is the decay rate into the waveguide, as can be understood in terms of absorption and stimulated emission of a photon as shown in Fig.~\ref{fig:schem}(b). Since the interaction is chiral, the delay is also proportional to the number of emitters $N$ in the waveguide. We also show that the pulse distortion encountered by an $n$-photon bound state scales as $n^{-6}$. Pulses of higher-number bound states therefore propagate with negligible distortion, i.e. much like a soliton. Moving beyond the few-photon--few-atom limit we obtain a simple description of photon propagation with a mesoscopic number of photons $n$ but a large number of atoms $N\gg 1$. Here, the number-dependent group delay breaks the  input pulse apart and produces a many-body ordered state of light. Finally, we show that the system approaches the classical limit for $n \gg 1$ and $n \gg N$, where our full quantum description captures the quantum-to-classical transition and reproduces the mean-field results of soliton propagation in self-induced transparency.

The effect discussed here has features similar to vacuum-induced transparency (VIT) where photons undergo a photon-number-dependent delay after interacting with atoms coupled to a cavity \cite{Nikoghosyan2010PRL, TanjiSuzuki2011Science}. A key difference is that the delay in VIT mainly depends on the total photon number inside the entire system, and not on the details of the pulse shape. The nonlinear effect we consider is spatially localized to the individual atoms and occurs for the simplest possible configuration of TLSs. This leads to a different spatiotemporal behaviour which we evaluate in a full multimode theory of the dynamics. Our theory features a full quantum many-body treatment of the system, where the photon time delay is examined by considering the many-body photonic scattering eigenstates.


We also point out that in mean-field theories solitons are known to be highly stable objects which are unaffected by external perturbations. Here we show that similar properties exist for few-photon bound states. We outline how one can conduct scattering experiments between photon bound states and individual photons. In the considered scattering experiments the bound state is deflected by the interaction but is otherwise unperturbed by it. Together the results obtained here demonstrate that the bound states should be considered as truly distinct physical entities emerging from the underlying light-matter interaction between photons and two-level emitters.

This manuscript is arranged as follows: in Section \ref{sec:model} we introduce the model for chiral wQED and outline various theoretical approaches for computing the dynamics through \ref{subsec:meanField} mean-field theory, \ref{subsec:eigenstates} the photon scattering eigenstates, and \ref{subsec:MPS} the matrix-product states (MPS). In Section \ref{sec:propagation} we compute how the input pulse propagates through the medium and compute the representation in terms of photon bound states. This is followed by Section \ref{sec:SITconnection} which shows that the many-body photon bound states can be used to construct the mean-field soliton solutions obtained in self-induced transparency. In Section \ref{sec:interactions} we show that photon bound states can undergo elastic scattering with individual photons modifying the delay of the bound state but otherwise leaving it unaltered, much like classical solitons. In Section~\ref{sec:outlook}, we show few-photon bound-state propagation can potentially be observed in state-of-the-art experiments with a few emitters, and we discuss potential future applications. Finally, we conclude in  Section~\ref{sec:conclusion}.

\section{Model} \label{sec:model}

We consider a system of $N$ two-level systems (TLS) chirally coupled coupled to a linearly dispersive one-dimensional bath of photons. Here, chiral coupling means that the TLSs couple only to right-propagating photons.  The Hamiltonian for this system ($\hbar =1$) is
\begin{equation}
\label{eq:Hamiltonian}
\hat{H} = -i \int dx \, \hat{a}^\dagger (x) \partial_x \hat{a}(x) + \sqrt{\Gamma}\sum_{i=1}^N  \left[\hat{\sigma}_i^- \hat{a}^\dagger (x_i) + \hat{\sigma}_i^+ \hat{a} (x_i)\right],
\end{equation}
where all integrals are over $\Re$, $\hat{\sigma}_i^\pm$ are Pauli operators for the $i$th TLS, $\hat{a}(x)$ ($\hat{a}^\dagger(x)$) is a photon annihilation (creation) operator at position $x$, the group velocity is set to unity $v_g=1$, and the energy has been renormalized to that of the TLS. In the limit of ideal chiral coupling, the system dynamics are not influenced by the positions of the TLSs $x_i$.  In this Hamiltonian the first term captures the free-propagation of photons in the waveguide with linear dispersion, while the next terms take into account the interaction between the photons and the emitters. In the absence of the interaction terms the photonic eigenstates of the Hamiltonian are plane waves with wavevector $k$ and frequency $\omega=k$. 
Equation (\ref{eq:Hamiltonian}) constitutes the typical scenario for chiral wQED \cite{Lodahl2017Nature}.  
In this manuscript, we are interested in describing the propagation of a multi-photon input field through the system. This goal can be achieved by exactly computing the scattering eigenstates of  Hamiltonian \eqref{eq:Hamiltonian}, as illustrated in Subsection \ref{subsec:eigenstates}.

In addition to computing the eigenstates of (\ref{eq:Hamiltonian}), we also introduce an equivalent formulation that is well-suited for using the MPS technique that we are going to introduce later in Subsection \ref{subsec:MPS}.
This approach is based on the observation that the transmitted field depends directly on the emitters' evolution. This can be seen by formally integrating the Heisenberg equation for the field operator $\hat{a}(x,t)$, that provides, within Born-Markov approximation, the following generalized input-output relation
for the transmitted field~\cite{Caneva2015NJP, Manzoni2017NCOM}
\begin{equation}
\label{eq:in_outR}
\hat{a}_{\rm out}(t)=\mathcal{E}_{\rm in}(t)+i\sum_j\sqrt{\Gamma} \hat\sigma_{j}^-(t),
\end{equation}
where we have defined $\hat{a}_{\rm out}(t)=\hat{a}(x^+_N,t)$ as the output field measured right after the last atom.
Note that, within this approach, we assume the input field $\mathcal{E}_{\rm in}(t)$ to  be a classical coherent field on resonance with the atomic transitions. With these assumptions the emitter dynamics driven by the input field is known to be described by a purely dissipative chiral master equation (ME) of the  form~\cite{Stannigel2012NJP, Pichler2015PRA} (see Supplemental Material (SM) for more information):
\begin{equation}\label{eq:MasterEq}
\dot \rho= -i \left( H_{\rm eff}\rho-\rho H^{\dagger}_{\rm eff}\right)+\Gamma\sum_{ij}\hat\sigma_{i}^-\rho \hat\sigma_{j}^+.
\end{equation}
Here,
\begin{equation}\label{eq:Heff}
H_{\rm eff}=-i\frac{\Gamma}{2}\sum_j\hat\sigma_{j}^+\hat\sigma_{j}^-+H_{\rm drive}-i\Gamma\sum_{l>j}\hat\sigma_{l}^+\hat\sigma_{j}^-,
\end{equation}
is the effective  Hamiltonian, which provides the non-Hermitian collective evolution of the emitters, while  the term $H_{\rm drive}=\sum_j\sqrt{\Gamma}\left[\mathcal{E}_{in}(t)\hat\sigma_{j}^++ \rm H.c.\right]$ gives the coupling of the emitters to  the input field.

The combination of Eqs.~\eqref{eq:in_outR} and \eqref{eq:MasterEq} provide a full description of the photon propagation through the chiral medium. In particular the spin dynamics can be efficiently solved by making use of an MPS ansatz~\cite{Verstraete2008ADPHYS, Schollwock2011Annals} as recently described in Ref.~\cite{Manzoni2017NCOM}. As will be shown in the following, this approach will allow us to fully explore the  limit of many photons and large atomic arrays, a scenario that is challenging to simulate with standard numerical techniques.

\subsection{Mean-field Theory and Self-induced Transparency} \label{subsec:meanField}

Before considering the full many-body dynamics of the Hamiltonian (\ref{eq:Hamiltonian}), we consider the system within the mean-field limit. We present this mean-field limit to contrast its predictions with the full many-body theory presented below.

The first treatment of (\ref{eq:Hamiltonian}) within mean-field theory dates back to the work on self-induced transparency (SIT) \cite{McCall1967PRL, McCall1969PR, Bullough1974OptoElectronics}. In these early experiments,  gasses of two-level atoms were excited by short intense laser pulses. Although the atoms are not ideally coupled to a single waveguide mode in such systems, the laser pulses are sufficiently short that decay channels to modes other than the laser mode can be neglected. Furthermore, both the weak coupling of the atoms to this mode and the high intensity of the laser means that one can consider the atoms as a spin continuum under the mean-field approximation where quantum correlations between the atoms and the light field can be neglected. Under these approximations, the equations of motion give the SIT equations
\begin{equation}\label{eq:MF}
\begin{split}
\left[ \frac{\partial}{\partial t}  + \frac{\partial}{\partial x} \right] a(x,t) = - i \sqrt{\Gamma}\sigma_-(x,t),\\
\frac{\partial}{\partial t} \sigma_-(x,t) = i \sqrt{\Gamma} \sigma_z(x,t)a(x,t),\\
\frac{\partial}{\partial t} \sigma_z(x,t) = 4 \sqrt{\Gamma} \operatorname{Im}{\left[a(x,t)\sigma_-^* (x,t) \right]}.
\end{split}
\end{equation}
Here, $a(x,t) = \langle \hat{a}(x,t) \rangle$, $\hat{\sigma}_{-/z} (x,t) = \sum_i \hat{\sigma}_i^{-/z}(t)\delta(x - x_i)$, and $\sigma_-(x,t) = \langle \hat{\sigma}_-(x,t) \rangle$, $\sigma_z(x,t) = \langle \hat{\sigma}_z(x,t) \rangle$ are the expectation values of the spin operators in the continuum limit. These nonlinear equations of motion have SIT soliton solutions. Following the treatment in \cite{Bullough1974OptoElectronics}, for a resonant pulse the field can be taken to be real, and one can map the equations of motion onto a nonlinear pendulum equation that can be solved exactly. This gives the fundamental soliton solution for the field
\begin{equation}\label{eq:soliton}
a(x,t) = \frac{\sqrt{\Gamma} \bar n}{2} \operatorname{sech}{\left[\frac{\Gamma \bar n}{2} \left(\frac{x}{V'} - t \right) \right]},
\end{equation}
where $\bar n$ is the number of photons in the field, (or more precisely, the total energy in the original SIT work). The pulse velocity within the medium in the laboratory frame is $V'=\bar n^2 \Gamma/(\bar n^2 \Gamma + 4 \nu)$, where $\nu$ is the gas density (see SM for details). Transforming to a frame comoving with the pulse in the absence of emitters, i.e. at velocity $v_g=1$, gives the relative velocity in the backwards direction $V = 4 \nu/(\bar n^2 \Gamma + 4 \nu)$. This corresponds to each emitter imparting a delay of $\tau_{\bar n} = 4/(\bar n^2 \Gamma)$ on the pulse. 

An important feature of the solitonic solution~\eqref{eq:soliton} is that the integrated Rabi frequency $\Omega = 2 \sqrt{\Gamma} \int dt a(x,t)$, which is proportional to the area under the pulse, is fixed by the relationship between the pulse amplitude and pulse width, and always evaluates to be  $2\pi$. This corresponds to a full Rabi cycle of complete excitation and subsequent deexcitation.  The SIT soliton therefore can be physically interpreted as a rapid excitation and de-excitation of the atoms, which suppresses spontaneous emission of the excited state and thus makes the medium transparent.

Inspired by the apparent dependence of velocity on photon number, it is interesting to ask whether this property extends to the few-photon limit, thus enabling, e.g., photon-number separation at the output. A full quantum treatment is necessary to answer this question, which we turn to in the following sections.

\subsection{Many-body Scattering Eigenstates} \label{subsec:eigenstates}

In contrast to the mean-field treatment we now consider the full many-body eigenstates. Since the Hamiltonian (\ref{eq:Hamiltonian}) preserves the combined number of atomic and photonic excitations, eigenstates with different numbers of excitations decouple. By computing the eigenstates in the one- and two-excitation subspaces, one can generalize the result to an arbitrary number of excitations. This technique is often referred to as Bethe's ansatz \cite{Bethe1931ZPhys} and is used to diagonalize a class of one-dimensional many-body Hamiltonians \cite{Thacker1981RMP}. In particular, it has previously been used to diagonalize the Hamiltonian in Eq.~(\ref{eq:Hamiltonian}) \cite{Rupasov1984JETP}. Since we are interested in the state that emerges after interaction with the TLSs, we are interested in the scattering eigenstates. These are the photon states that interact with the TLSs and emerge unchanged apart from an overall transmission coefficient. The $n$-body scattering eigenstates have the form
\begin{widetext}
\begin{equation}
\label{eq:MBScattEig}
| S_\mathbf k \rangle_{n} = C_{\mathbf k, n, S} \frac{1}{\sqrt{n!}} \int d^n \mathbf{x} \,  \hat{\mathbf{a}}^\dagger (\mathbf{x}) | 0 \rangle \prod_{l<m}^n \left[ k_l - k_m + i \Gamma \operatorname{sgn}{(x_{l} - x_{m})} \right] \prod_{j=1}^n e^{ i k_j x_{j}} + \leftrightarrow,
\end{equation}
\end{widetext}
where $C_{\mathbf k, n, S}$ is a normalization constant which varies with wavevector $\mathbf k$, excitation number $n$, and the type of eigenstates $S$, where $S$ labels different states as explained below; $\hat{\mathbf{a}}^\dagger (\mathbf{x}) = \hat{a}^\dagger (x_1) \hat{a}^\dagger (x_2) \ldots \hat{a}^\dagger (x_n)$; $d^n \mathbf x = dx_1 dx_2 \ldots dx_n$; and $\leftrightarrow$ indicates summing over all $n!$ permutations of $x_i$ to symmetrize the wavefunction. The energy of the eigenstates is $E=\sum_i^n k_i$. Upon scattering off all $N$ emitters, the eigenstates are multiplied by the eigenvalue $t_{\mathbf k}^N=\prod_{j=1}^n t_{k_j}^N$, where $t_k = (k - i \Gamma/2)/(k + i \Gamma/2)$. Since, by assumption, the system does not contain any dissipation and is chiral, all the transmission coefficients have $|t_\mathbf k|=1$. This means that the transmission coefficients simply multiply the eigenstate by a phase. For example, a single photon on resonance undergoes a $\pi$-phase shift with $t_0=-1$. Importantly, the phase that is imparted on the eigenstate varies with $\mathbf k$, i.e., the TLSs introduce dispersion to the system. As we soon show, different types of eigenstates also accumulate different phases. We note that the output states are described using the position coordinates $\mathbf x$. This is equivalent to using the time variable $-t$ in (\ref{eq:in_outR}) as we have set the group velocity to unity. 

In addition to different eigenstates for different values of $n$, there are also different possible types of eigenstates within different excitation-number manifolds. In the single excitation subspace there is only one type of eigenstate and it is characterized by a real wavenumber $k$ and fully extended in space. For $n=2$, the wavenumbers $k_1$ and $k_2$ can either both be real which gives rise to a fully extended solution of the form (\ref{eq:MBScattEig}). They can also assume complex values $k_1 = E/2 + i \Gamma/2$ and $k_2 = E/2 - i \Gamma/2$ which are called strings \cite{Thacker1981RMP}. These values give additional valid solutions. Since the wavevectors can form complex conjugate pairs, the eigenstates become localized in a relative coordinate while being fully extended within the center-of-mass coordinate. This localization is associated with the formation of {\it photon bound states} that manifest in the bunching of two photons which travel together during their propagation \cite{Shen2007PRA, Zheng2011PRL}.  

For $n>2$ the $m$-body string ($m \leq n$) has the wavevector $k_j = K/m - i\left[ m+1 - 2j\right]\Gamma/2$ with $j=1,2,\ldots,m$ where $K$ is the total energy of the $m$-string. In total, the $n$-photon manifold has $p(n)$ string combinations where $p(n)$ is the number of partitions of $n$. For example, for three photons, there is a completely extended scattering eigenstate, a completely bound eigenstate, and a hybrid state with two bound photons and one extended photon. The different string combinations give the different types of scattering eigenstates $S$ by substituting the complex wavevectors into (\ref{eq:MBScattEig}). The transmission coefficient is also obtained by substituting the complex wavevectors into the expression for $t_{\mathbf k}$.

For $n$ photons, the $n$-string state is a fully localized bound state with energy $E$,
\begin{equation}
\label{eq:MBBound}
| B_E \rangle_n  = \frac{C_{n, B}}{\sqrt{n!}}\int d^n x \, \hat{\mathbf a}^\dagger(\mathbf x) | 0 \rangle
 e^{i \frac{E}{n} \sum_j x_j -\frac{\Gamma}{2} \sum_{i<j}|x_i - x_j|},
\end{equation}
where $C_{n,B} = \sqrt{\Gamma^{n-1} (n-1)!/(2 \pi n)}$. The transmission coefficient for the $n$-photon bound state is then,
\begin{equation}\label{eq:transmission}
\begin{split}
t_{E,n} = \frac{E - i \Gamma n^2/2}{E + i \Gamma n^2/2}.
\end{split}
\end{equation}
Importantly, the phase of the transmission coefficient varies with $n$, i.e., the system has a photon-number-dependent dispersion. 

With all the eigenstates at hand, the scattering matrix for interacting with all $N$ emitters in the $n$-photon manifold can be formally written as
\begin{equation}
\begin{split}
\label{eq:ScattMat}
\hat{S}_n^N = \sum_{S} \sum_{\mathbf k} t_{\mathbf k}^N | S_{\mathbf k} \rangle_{n} {}_n \langle S_{\mathbf k} |,
\end{split}
\end{equation}
where the sum over $S$ is a sum over the different string combinations of the $n$-photon manifold. We note that the eigenstates are orthogonal, thus the scattering matrix for $N$ emitters simply requires taking the eigenvalue to the $N$-th power. Since the number of string combinations increases as $p(n)$, the number of terms in the sum increases exponentially for large $n$ \cite{Rupasov1984JETP, Shen2015PRA}. In this manuscript we compute the full output states for up to $n=3$ using this formalism. We note that a formalism exists where one does not have to sum over string combinations \cite{Yudson1985JETP}. Nevertheless the form used here is particularly insightful as it gives direct access to the number-dependent transmission coefficient which, as we show, plays a central role in understanding the many-body pulse propagation. 

\begin{figure*}[!t]
\includegraphics[width=1.8\columnwidth]{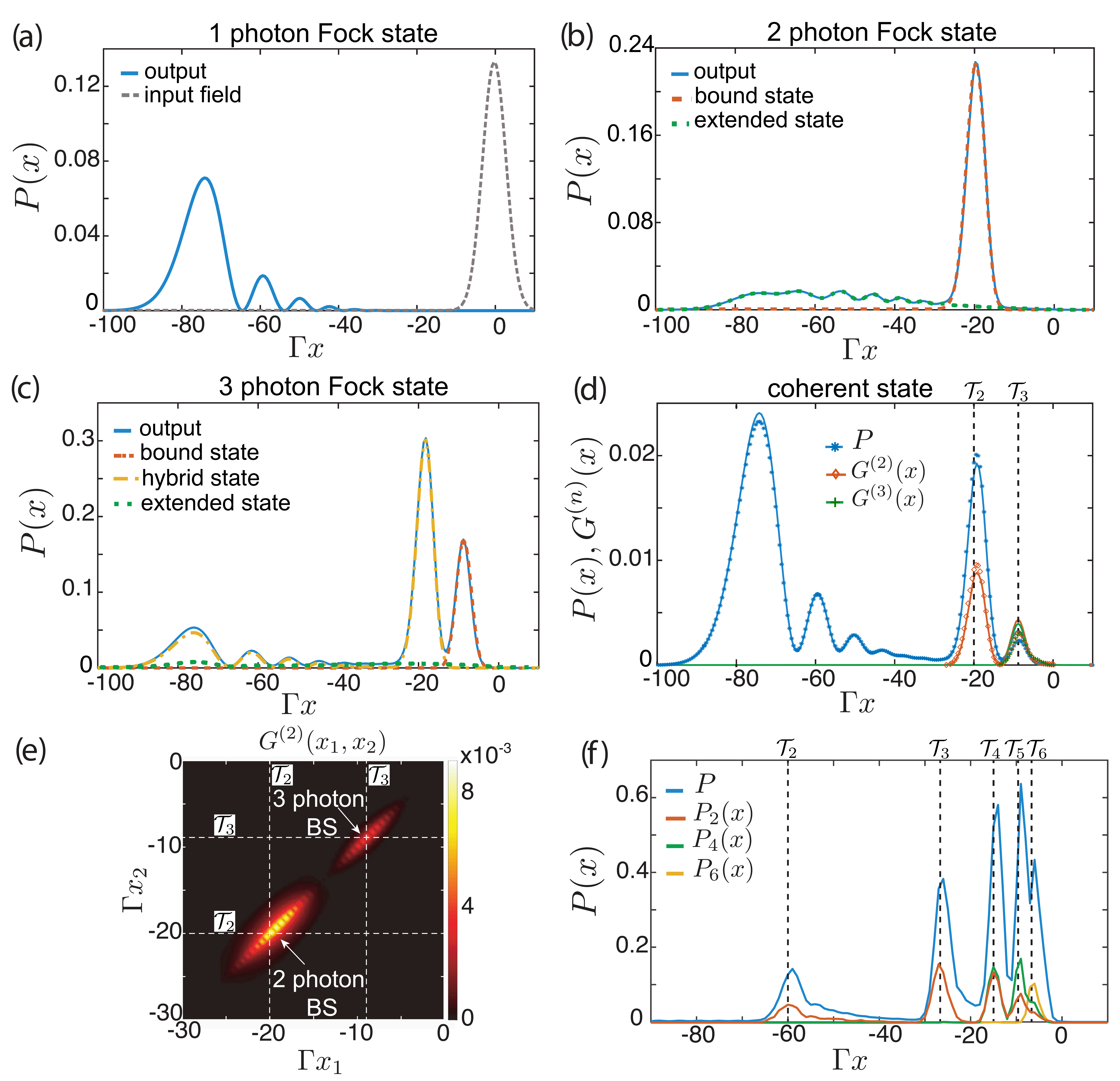}
\caption{\label{fig:propagation} Transmitted power $P(x)$ (solid curves) for (a) one- (b) two-, and (c) three-photon transport through $N=20$ TLSs for a resonant pulse with $\Gamma \sigma=3\sqrt{2}$ computed using the photon scattering theory. In (b) and (c) dashed lines show the contributions from the different eigenstates. The input mode in all panels is the Gaussian curve shown in (a). (d) Transmitted power and correlation functions for a coherent state with $\bar n=0.5$. Solid lines show photon scattering results truncating at three photons while marker points show MPS calculations performed by solving the ME with a Runge-Kutta algorithm and using maximum bond dimension $D_{\rm max}=150$. (e) Second-order correlation function $G^{(2)}(x_1,x_2)$  for the transmitted coherent state computed using photon scattering theory. The two- and three-photon bound state contributions appear as peaks that are localized in the relative coordinate. (f) MPS calculation for input coherent state with $\bar n =8$, $N=60$ emitters and $\Gamma \sigma=2\sqrt{2}$. Dashed vertical lines show $N$ times the Wigner delay ${\cal T}_n = N \tau_n = 4N/(\Gamma n^2)$. With $P_n(x)$ we indicate the contribution to the intensity coming from the emission of $n$ photons within the space bin $\Delta x=1$. To perform this calculation we used a quantum trajectories MPS algorithm fixing the maximum bond dimension to $D_{\rm max}=40$.}
\end{figure*}

\subsection{MPS Ansatz} \label{subsec:MPS}

In order to study the dynamics for  stronger input pulses ($n>3$)  than the one computed with the $S$-matrix formalism,  we solve equations~\eqref{eq:in_outR} and \eqref{eq:MasterEq} using an MPS ansatz. Specifically, the system evolution can be solved either by directly solving the ME~\eqref{eq:MasterEq}~\cite{Bienias2018arXiv} (method used for Fig.~\ref{fig:propagation}(d) and Fig.~\ref{fig:towardsSIT}) or by using a quantum trajectories algorithm where the state of the system evolves under the effective Hamiltonian~\eqref{eq:Heff} and stochastically experiences quantum jumps~\cite{Manzoni2017NCOM} (method used for Fig.~\ref{fig:propagation}(f)). In both cases an MPS representation is applied either to the quantum state or to a vectorized form of the density matrix. Here for simplicity we limit the discussion to the former while the latter is discussed in the Supplemental Material. 

The MPS ansatz consists of reshaping the generic quantum state $|\phi \rangle=\sum_{i_1,..i_N}\psi_{i_1,i_2,..i_N}|i_1,i_2,..i_N\rangle$ (with $i_j\in \{g,e\}$) into a matrix-product state of the form:
\begin{equation}\label{eq:MPS ansatz}
|\phi\rangle=\sum_{i_1,\ldots, i_N}A_{i_1}A_{i_2}\ldots A_{i_N}|i_1,i_2,\ldots, i_N\rangle,
\end{equation}
where, for each specific set of physical indices $\{ i_1,i_2,\ldots, i_N\}$, the product of the $A_{ i_j}$ matrices gives back the state coefficient $\psi_{ i_1, i_2,\ldots, i_N}$. Each matrix $A_{i_j}$ has dimension $D_{j-1}\times D_{j}$ known as the bond dimension and finite edge boundary conditions are assumed by imposing $D_{1}=1$ and $D_{N}=1$. The bond dimension reflects the entanglement entropy. For instance, if $D_{j}=1$ for all $j$, the matrices $A_{i_j}$ are scalars and the state \eqref{eq:MPS ansatz} reduces to a product state with no entanglement. For arbitrary states the bond dimension grows exponentially with the number of particles. The advantage of the MPS ansatz is that, in many physical scenarios as the one considered here, the entanglement grows slowly with the system size allowing an efficient description of the state in terms of a smaller bond dimension~\cite{Schollwock2011Annals}.  An important figure of merit of the efficiency of the MPS ansatz is given by $D_{\rm max}$, the maximum bond dimension, that is needed to faithfully represent the system during the entire time evolution (see SM for more information). We will make use of this quantity in Sec.~\ref{sec:SITconnection} to quantify the amount of many-body correlations present in the system.

\section{Many-body Pulse Propagation} \label{sec:propagation}

We are interested in studying multi-photon propagation through the chirally coupled array. Here, we consider coherent and Fock input states with mode creation operator $\hat{a}^\dagger_{\rm in} = \int dx \mathcal{E}(x)\hat{a}^\dagger(x)$, and we evaluate the transmitted field with the two methods described in the previous section. In particular, for the exact solution we compute the transmitted photon state using the eigenstates for up to three photons while for higher excitations we make use of the  MPS ansatz. In Fig. \ref{fig:propagation} we consider the propagation of a  Gaussian photonic mode with amplitude $\mathcal{E}(x)=e^{i k_0 x - x^2/(2 \sigma^2)}/(\sqrt{\sigma}\pi^{1/4})$ where, throughout the manuscript, resonant pulses are considered $k_0=0$.

Figure \ref{fig:propagation} (a)-(c) shows the power $P(x) = \langle \hat{a}^\dagger (x) \hat{a}(x) \rangle$ for one-, two-, and three-photon Fock states after propagating through $N=20$ TLSs. Here, $x=0$ is chosen to be the reference frame of the pulse propagating in the absence of emitters. A pulsewidth of $\Gamma \sigma = 3 \sqrt{2}$ is chosen to have appreciable overlap with all the different types of scattering eigenstates while remaining sufficiently narrow to observe the photon-number-dependent velocities. The magnitude of the overlap of an input Gaussian pulse with two- and three-photon bound states versus Gaussian pulsewidth $\sigma$ is shown in the Supplemental Material.   In Fig.~\ref{fig:propagation}(b) we see that the two-photon bound state comes out earlier than the extended state. The two-photon bound state thus clearly propagates with a faster velocity than the extended state. The bound state also undergoes significantly less broadening and distortion. For the three-photon transport in Fig.~\ref{fig:propagation}(c) again the extended state is distorted and delayed, while the three-photon bound state has significantly less distortion and delay. In the three-photon manifold there is also a string combination that forms a hybrid state where one photon is completely extended while the other two are bound. The evolution of this state is determined by the individual components of the state: the two bound photons propagate in a similar manner to the two-photon bound state in Fig.~\ref{fig:propagation}(b), while the extended photon propagates like a single photon. This separation of the propagation of the bound and extended parts of this state can be shown explicitly in the long pulse limit $\sigma \Gamma \gg 1$ (see SM). This is significant because it implies that in order to understand the pulse evolution one does not have to understand the behaviour of all the different string combinations $S$. Rather, one can focus on simply understanding the behaviour of the bound states. On the other hand, for short pulses this separation is not completely true because one needs to include the effect of interactions between the components, see Sec. \ref{sec:interactions}.

\subsection{Evolution of bound states}  

To better quantify the difference in propagation observed above, let us compute the pulse evolution in the center-of-mass coordinate. This can be done by using the form of the $n$-photon bound state and its transmission coefficient given in Eqs.~\eqref{eq:MBBound} and \eqref{eq:transmission}.
 Within the $n$-photon manifold the projection of an input Gaussian state on the bound state is ${}_n\langle B_{E} | \textrm{in} \rangle = c_n e^{-(E - n k_0)^2 \sigma^2/(2n)}$, where $c_n$ is a constant in $E$. Here it is convenient to use Jacobi coordinates $x_c = \sum_j^n x_j/n$, $x^J_i = \sum_{j=1}^i x_j/i - x_{i+1}$, where $i\in \{1,2,\ldots, n-1 \}$. The resulting bound-state contribution is then  
\begin{equation}
\begin{split}
|\textrm{out} \rangle_{n,\textrm{bound}} =& \frac{c_n}{\sqrt{n!}} \int d^n x \, \hat{\mathbf a}^\dagger(\mathbf x) | 0 \rangle e^{-\frac{\Gamma}{2}g(\mathbf x^J)} \\
&\times \int dE \, t_{E,n}^N e^{i E x_c -E^2 \sigma^2/(2n)}, 
\end{split}
\end{equation}
where $g(\mathbf x^J)$ is the exponent $\sum_{i<j}|x_i - x_j|$ written in Jacobi coordinates. The center-of-mass evolution is clearly determined by the second integral. This now has the standard form of Gaussian pulse propagation through a linear dispersive medium. Defining $t_{E,n}^N \equiv e^{i N \phi(E)}$, the first to third derivatives of $\phi(E)$ give respectively, the delay, broadening, and distortion that the pulse undergoes per emitter. The delay per emitter is $\tau_n(k_0) = \Gamma/(k_0^2 + n^2 \Gamma^2/4)$. This is largest for a resonant pulse ($k_0=0$), where 
\begin{equation}
\label{eq:timeDelay}
\tau_n = \frac{4}{n^2\Gamma}.
\end{equation}
This gives the Wigner delay imparted on an $n$-photon bound state by a single emitter in wQED: the photons propagate with a number-dependent velocity. 

By taking higher-order derivatives we also compute the pulse broadening $b(k_0) = -32 k_0 \Gamma/[n (4k_0^2 + n^2 \Gamma^2)^2]$, which is zero on resonance. The third-order pulse distortion term on resonance is $d = -32/(n^6 \Gamma^3)$. The pulse distortion is thus drastically reduced for higher-order bound states. This indicates that many-photon bound states suffer negligible pulse distortion while propagating through the array of nonlinear and dispersive atoms. 

In order to verify that indeed the physics of the bound states dominates the wave-packet evolution we also compute the evolution of a coherent input pulse as shown in Figure \ref{fig:propagation}(d). Here, the pulse width $\Gamma \sigma = 3 \sqrt{2}$ is the same as in Figs. \ref{fig:propagation}(a)-(c), while the average photon number in the pulse is $\bar n = 0.5$. We compute the output both by truncating the coherent state to three photons, and solving exactly using Eq.~(\ref{eq:ScattMat}), or by solving Eqs.~\eqref{eq:in_outR} and\eqref{eq:MasterEq} with the MPS algorithm. The evolution of the bound states is seen in the position of the peaks of the power distribution $\langle \hat{a}^\dagger(x) \hat{a}(x) \rangle$ as well as in the difference between the power and the $m$-th order correlation functions $G^{(m)}(x)=\langle \left[ \hat{a}^\dagger(x)\right]^m\left[\hat{a}(x)\right]^m \rangle$. 

The localized nature of the bound states in the relative coordinates is shown in Fig.~\ref{fig:propagation}(e) where we show the two-point second-order correlation function $G^{(2)}(x_1,x_2)=\langle \hat{a}^\dagger(x_2)  \hat{a}^\dagger(x_1)\hat{a}(x_1) \hat{a}(x_2)\rangle$.  Here the photons tend to localize around the diagonal at small values of the relative coordinate $x_1 - x_2$, while they are delocalized about the center-of-mass coordinate $(x_1+x_2)/2$. This reflects that the photons are tightly bound together in the bound state, but the bound state itself is free to propagate. While exact analytical calculations become infeasible for $\bar{n}\gg 1$, the validity of our arguments and the importance of the bound states can still be seen in MPS simulations. For example, in Fig.~\ref{fig:propagation}(f), we calculate the transmitted power for an input coherent state $\bar n=8$ and $N=60$. Here, for this particular system length, the photon-number-dependent Wigner delay clearly manifests itself as separate peaks for up to six-photon bound states. 

The low-distortion propagation of the bound states can be intuitively explained by returning to the simple schematic shown in Fig.~\ref{fig:schem}(b). When a multi-photon bound state propagates trough the atomic array one of the photons in the wave packet can be absorbed and re-emitted by the atom. This process occurs on a time scale ruled by the inverse of the photon-number dependent stimulated emission rate that coincides with the bound-state packet width, $\Delta t_n \sim 1/(\Gamma n)$, allowing the pulse to preserve its shape. This continuous absorption and re-emission of photons during the bound-state propagation leads to a time delay of one out of $n$ photons by an amount $4/(\Gamma n)$, leading to the group delay in Eq.~(\ref{eq:timeDelay}).

\subsection{Influence of imperfections}

We have shown that the hallmark of many-body photon propagation through an ensemble of quantum emitters in chiral wQED is the number-dependent velocity of the photon bound states. Here we analyse the influence of imperfections such as losses, imperfect chirality, and inhomogeneous broadening on this propagation. We first consider the influence of losses, where each emitter couples to an additional decay channel out of the waveguide with a rate $\Gamma_0$. It is possible to obtain an analytic criterion for when these losses can be neglected. The form of the transmission coefficient of the bound state can be obtained in the presence of loss by mapping the total energy to a complex energy using the replacement $E \rightarrow E + i n \Gamma_0/2$. The reduction in probability of the output state then implies that the state can undergo one or more quantum jumps. If the probability remains close to unity the output state is only weakly affected. Defining the efficiency or $\beta$-factor as $\beta = \Gamma/\Gamma_{\rm tot}$, where $\Gamma_{\rm tot} = \Gamma + \Gamma_0$, the transmission coefficient (\ref{eq:transmission}) in the presence of loss is  
\begin{equation}
\label{eq:tBoundLoss}
t_{E,n} = \frac{E + i n \Gamma_{\rm tot} [1-\beta (1+n)]/2}{E + i n \Gamma_{\rm tot} [1-\beta (1-n)]/2}.
\end{equation}
After scattering off all $N$ emitters the magnitude of the resulting state is $|t_{E,n}|^N$. For small imperfections $1-\beta \ll 1$, this gives $|t_{E,n}|^N = 1- 2N(1-\beta)/n + \mathcal{O}\left((1-\beta)^2 \right)$, where the notation $\mathcal{O}(M)$ indicates term of order $M$ and higher. This means that a sufficient condition for neglecting losses is $N(1-\beta)/n \ll 1$. This implies that losses have a reduced influence on higher-order bound states. If this condition is not met, there is a sizable probability that one or more of the photons in a photon bound state is lost. 

If a photon is lost at one point along the ensemble the remaining photons propagate through the rest of the atoms with a different effective group velocity and dispersion.  In addition to a reduced ampltitude, the fact that the remaining transmitted photons have effectively propagated with a mix of velocities and dispersions causes the peaks associated with the different bound states in Fig.~\ref{fig:propagation} to broaden and eventually overlap (see SM).


In addition to photon loss, we note that a non-zero value of $\Gamma_0 \neq 0$ also affects the delay $\tau_n$. This is also computed analytically, $\tau_n = 4\beta^3/\{\Gamma [\beta(2+\beta(n^2-1))-1] \} =4/(\Gamma n^2) - 4(1-\beta)/(\Gamma n^2) + {\mathcal O}((1-\beta)^2)$. Values where $\tau_n$ diverges occur when $|t_{E,n}|$ approaches zero, i.e., no light is transmitted.

In the limit of large losses the photon bound states suffer exponential damping. For a steady-state input field, this was considered for up to two photons in Ref.~\cite{Mahmoodian2018PRL} where it was shown that the output shows strong photon bunching. This bunching, however arises from the extended states and does not reflect the bound-state dynamics.  Even with finite pulse durations it is likely that the bound-state dynamics will be hard to discern in this limit. 

The introduction of imperfect chirality also influences the nature of pulse propagation through the ensemble. This occurs when, in addition to coupling to the forward-propagating mode, each atom also couples to the backward propagating mode with a rate $\Gamma_L$. The influence of this cannot be considered analytically in a straightforward manner. We have performed numerical MPS calculations (see SM) for $N=20$ emitters. When $\Gamma_L=0.05\Gamma$, the shape of the output state remains qualitatively unchanged. However, when $\Gamma_L=0.2\Gamma$ the output state is completely distorted. These computations indicate that the effect observed in the ideal chiral case is robust to imperfect chirality, provided that the imperfections are not too large.

We also consider the influence of inhomogeneous broadening in the two-level systems. The two-level atoms are considered to have a normally distributed resonance frequency with dimensionless standard deviation  $\varsigma/\Gamma$. This affects the transmission coefficient of the bound states, which most notably affects the pulse delay. An expression for the mean pulse delay can be computed analytically, and for small broadening, gives to leading order   
\begin{equation}
\label{eq:tBoundLoss}
\langle \tau_n \rangle = \tau_n \left[ 1- \frac{4 \varsigma^2}{n^2 \Gamma^2}  + \mathcal{O}\left(  \frac{\varsigma^4}{\Gamma^4} \right) \right].
\end{equation}
The reduction in the delay imparted by each emitter therefore scales quadratically in $\varsigma/(n \Gamma)$. This therefore has a reduced impact on higher-order bound states provided that the broadening is limited $\lesssim \Gamma$. We note that the influence of inhomogeneous broadening can be compensated by introducing more emitters.


\section{Connection to the SIT soliton} \label{sec:SITconnection}

In the previous sections we have shown how the Hamiltonian~\eqref{eq:Hamiltonian} leads to the SIT solitonic solutions in the mean-field limit, and that the full quantum mechanical treatment of this Hamiltonian predicts correlation-ordered photon propagation. In this section we aim to bridge the gap between these two regimes: first we show that indeed the many-body theory reduces to the mean-field result in the limit of large photon number. Secondly, we derive the quantum corrections to the mean-field results which become relevant when both the number of photons $n$ and the number of emitters $N$ are large. Finally, we push the numerical simulations to the many-photon limit to verify the analytical predictions.

Let us consider a wave packet composed of a linear combination of many-body bound states. This is expressed with the ansatz
\begin{equation}
\label{eq:boundAnsatz}
|\psi\rangle =  \sum_n \int dE c_n(E) | B_{E} \rangle_n.
\end{equation}
We later show that this is the expected form of the state for a high-power coherent input with a large spectral width. Unlike the previous sections where we selected an input pulse and propagated it through the medium, here we are simply considering a linear combination of bound eigenstates and compute observables for this state. A localized function $c_n(E)$ ensures that the bound state ansatz is localized in the center-of-mass coordinate.

Such a state can be probed by measuring either the field $\langle \psi | \hat{a} (x)| \psi \rangle$ or the $m$-th normally ordered observable, which we consider in the center-of-mass coordinate $\langle \psi | (\hat{a}^\dagger(x))^m (\hat{a}(x))^m | \psi \rangle$. We compute these observables in the limit where the average photon number is large $\bar{n} \gg 1$, the pulses are spectrally broad $\sigma \ll 1/\Gamma$ and the order of the correlation function is much less than the photon number $m \ll n$. These give (see SM for full calculations),
\begin{align}
\label{eq:MFfield}
\langle \psi | \hat{a} (x)| \psi \rangle &= \frac{\bar{n} \sqrt{\Gamma}}{2}\operatorname{sech}{\left(\frac{\bar{n} \Gamma x}{2}\right)},\\
\label{eq:MFmthOrder}
\langle \psi | (\hat{a}(x)^\dagger)^m (\hat{a}(x))^m | \psi \rangle &= \left[ \frac{\bar n \sqrt{\Gamma}}{2}\operatorname{sech}{\left(\frac{\bar{n} \Gamma x}{2}\right)} \right]^{2m},
\end{align}
where $\bar{n}$ is the average number of photons in the pulse. These observables reproduce the fundamental soliton solution of SIT \cite{McCall1969PR, Bullough1974OptoElectronics, Yulin_2018} given in Eq.~\eqref{eq:soliton}. Self-induced transparency is thus the classical limit of the photon bound state when the photon number becomes large, or conversely photon bound states are simply the quantum limit of a soliton, a {\it quantum soliton}. Just like the SIT solitons, the integrated Rabi frequency of the pulse is $\Omega = 2 \sqrt{\Gamma} \int dx \langle \hat{a}(x) \rangle = 2 \pi$, i.e., the intense pulse of light rapidly excites and de-excites the emitters resulting in a $2\pi$ Rabi oscillation. We note that, the expressions in Eqs.~(\ref{eq:MFfield})-(\ref{eq:MFmthOrder}) scale with $\Gamma$, which is the coupling to the one-dimensional continuum. These expressions are therefore unchanged in the presence of coupling to an external reservoir. This implies that, provided the ensemble does not act like a Bragg mirror, SIT is expected to occur even in the presence of backscattering in accordance with mean-field results.

\begin{figure*}[!t]
\includegraphics[width=1.3\columnwidth]{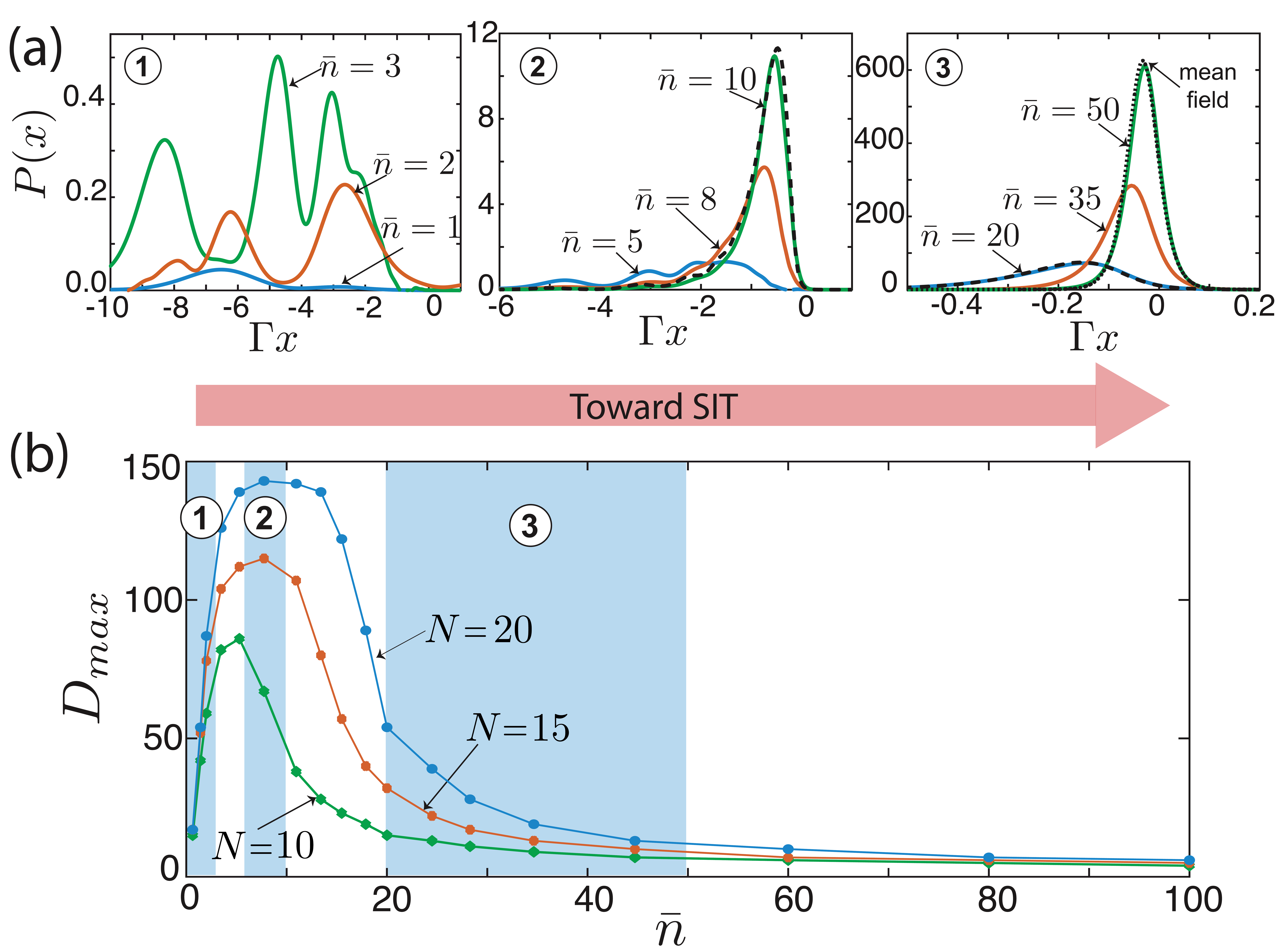}
\caption{\label{fig:towardsSIT} (a) Output power for $N=20$ atoms for coherent input pulses with a solitonic shape (see text) and different mean photon number $\bar n$. The results are computed by directly solving the ME~\eqref{eq:MasterEq} (solid lines) with the MPS algorithm where we fixed $D_{\rm max}=150$. In the second panel we include the results given in the many-body limit (\ref{eq:SITbeyondMF}) (dashed lines), and in the third the mean-field solitonic ansatz (dotted line). We note that the $x$-axis is limited to focus on larger photon number components. (b) Maximum  bond dimensions $D_{\rm max}$ of the MPS calculations required to obtain a tolerance less than $10^{-4}$ versus input power $\bar n$ for different TLS number $N$. The blue shading highlights the regions for which panels (1)-(3) in (a) show the output power.}
\end{figure*}

\subsection{Beyond mean-field theory}

As a lowest-order approximation, the variation in photon number $n$ making up the pulse (\ref{eq:boundAnsatz}) can be ignored, and the state will simply propagate at a reduced speed dictated by the mean photon number. This leads to the SIT soliton prediction, that simply maps $x \rightarrow x + 4N/(\bar{n}^2 \Gamma)$. However, for a coherent state, the uncertainty in the photon number scales as $\Delta n \sim \sqrt{\bar{n}}$, which leads to a gradual broadening of the pulse due to the different photon number components accumulating different time delays. This can become significant for a sufficiently large number of emitters $N$. This broadening is not captured by the mean-field theory. The breakdown of the mean-field theory therefore occurs when the difference in the delay of the $\bar n$ and the $\bar n + \Delta n$ photon bound states becomes on the order of the width of the $\bar n$ photon bound state. This gives $({\cal T}_{\bar n+ \Delta n}-{\cal T}_{n})/\Delta t_n \sim N/n^{3/2}$. This means that when $N/n^{3/2}\gtrsim 1$ the  mean-field theory breaks down even for large input photon number $\bar n$. This provides the boundary between mean-field theory and genuine quantum many-body dynamics.

When $\bar n \gg 1$ and $N \gtrsim n^{3/2}$, one can consider a wavefunction composed of bound states with a number distribution given by a coherent state. In the limit where mean-field theory breaks down, one must explicitly consider the Fock-state-dependent delay. In this limit the expression for the field and the power give (see SM),
\begin{equation}
\begin{split}
\label{eq:SITbeyondMF}
\langle \hat{a}(x) \rangle &= e^{-|\alpha|^2}\sum_n \frac{\alpha^{2n}}{n!} \frac{n \sqrt{\Gamma}}{2}\operatorname{sech}{\left[\frac{n\Gamma}{2} \left(x + \frac{4 N}{n^2 \Gamma}\right)\right]},\\
\langle \hat{a}^\dagger(x) \hat{a}(x) \rangle &= e^{-|\alpha|^2}\sum_n \frac{\alpha^{2n}}{n!} \frac{n^2 \Gamma}{4}\operatorname{sech}{\left[\frac{n\Gamma}{2} \left(x + \frac{4 N}{n^2 \Gamma}\right)\right]}^2,
\end{split}
\end{equation}
with equivalent expressions for higher-order correlation functions. Here, $\alpha = \sqrt{\bar n}$ is the coherent field amplitude, which is assumed real. We note that similar expressions exist for the bound state of the nonlinear Schr\"{o}dinger equation with an attractive interaction \cite{Lai1987PRAI, Lai1987PRAII}.

Equations (\ref{eq:SITbeyondMF}) provide a simple description of the observables of a quantum many-body state of light. In order to investigate the full transition from the multi-photon bound-state propagation to the formation of the SIT solitons we make use again of the master equation simulation. In Fig.~\ref{fig:towardsSIT}(a) we plot the transmitted power  of input pulses with $\mathcal{E}_{in}(x)=\frac{\bar{n} \sqrt{\Gamma}}{2}\operatorname{sech}{\left(\frac{\bar{n} \Gamma x}{2}\right)}$ for different amplitude strength $\bar n$. This pulse shape is chosen such that its electric field matches the SIT criterion. In the first box we see again the formation of the bound state peaks which tend to reduce their time delay as the input power is increased. For intermediate input pulses (second box) the bound states with a large number of photons get more populated, and they accumulate toward a single peak as the difference in delay times for large photon numbers become less distinguishable. In this regime the transmitted power starts to be well-described by Eq.~\eqref{eq:SITbeyondMF}. For even higher input power (third box) the individual bound states are no longer recognizable and a solitonic pulse well-described by mean-field theory emerges. These results show how the SIT solitons emerge from a superposition of photon bound states that, in  the limit of few photons, can be indeed interpreted as {\it quantum solitons}. On the other hand it is important to emphasize the difference in physical effects. While the SIT solitons can be fully described by a mean-field semi-classical treatment, the formation of distinct bound-state peaks is characterized by a highly-correlated state of light and represents the break down of the mean-field solution due to quantum effects.

Within the MPS ansatz, one natural way to characterize the amount of correlations in the system is to allow the maximum truncated bond dimension $D_{max}$ to vary, and to record the value $D_{max}^{th}(N,\bar{n})$ at which the truncation error exceeds some acceptable threshold value (see SM). In Fig.~\ref{fig:towardsSIT}(b), we show this quantity as a function of the solitonic input pulse strength $\bar n$. We see that the breakdown of the mean-field description occurs in the  regime where bound state formation occurs and the amount of correlation in the system is high (large values of $D_{max}$), while in the limit of large $\bar n$ the bond dimension tends to shrink approaching the mean-field limit $(D_{max}^{th}=1)$.

\section{Soliton Interactions} \label{sec:interactions}

\begin{figure}[!t]
\includegraphics[width=\columnwidth]{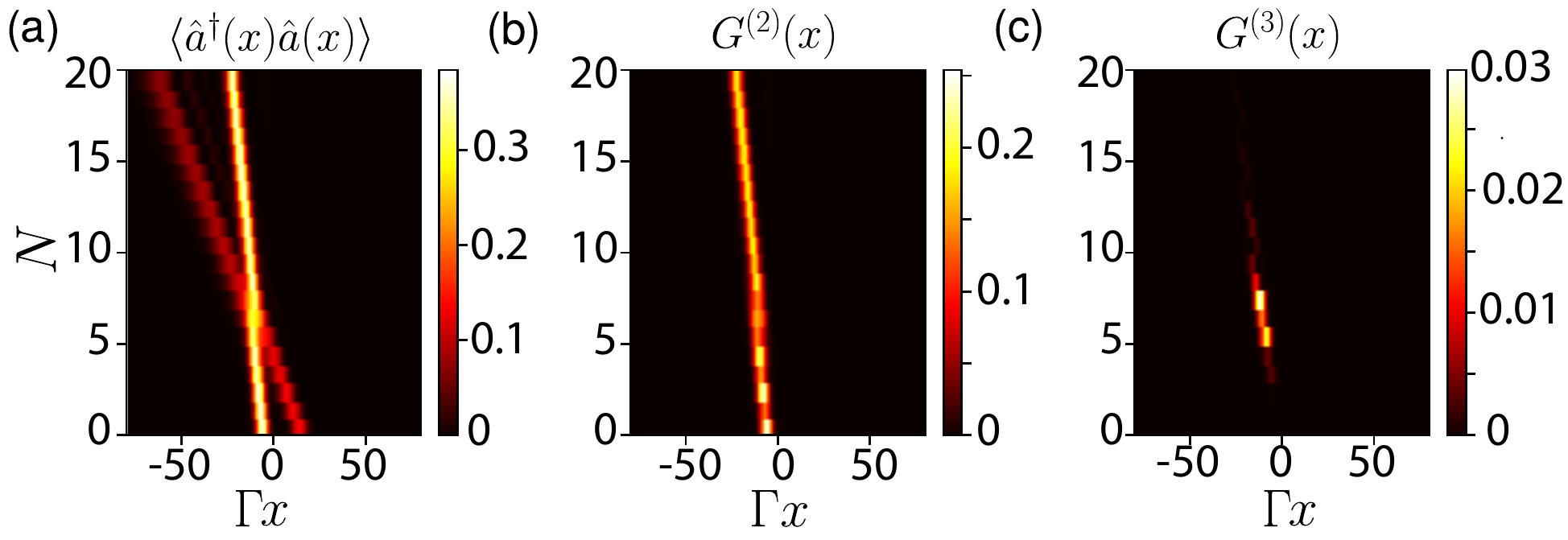}
\caption{\label{fig:interactions} Evolution of the state $|\textrm{in} \rangle$ (see text) demonstrating the interaction between a two-photon bound state centered at $a_2=-10/\Gamma$ and a single photon centered at $a_1=15/\Gamma$ both with width $\sigma \Gamma = 3\sqrt{2}$. The output observables (a) $\langle \hat{a}^\dagger (x) \hat{a}(x) \rangle$, (b) $G^{(2)}(x)$, and (c) $G^{(3)}(x)$ are plotted versus the number of emitters $N$.}
\end{figure}

So far we have characterized the propagation of light through ideal media and shown that it can be understood in terms of the photon bound states. To fully characterize and understand these objects it is important to also investigate their interactions and robustness to disturbances. To this end, we now make a preliminary investigation of a scattering experiment between a single photon and a two-photon bound state. We consider the input state $|\textrm{in} \rangle = C \int d^3 \mathbf x \, \mathbf a ^\dagger (\mathbf x) | 0 \rangle \phi(x_1,x_2,x_3)$ with $C$ being a constant and  
\begin{equation}
\begin{split}
\phi(x_1,x_2,x_3) = &e^{-(x_1-a_1)^2/(2\sigma^2)} e^{-(x_2+x_3 - a_2)^2/(4\sigma^2)}\\
\times &e^{-\Gamma/2|x_3-x_2|} + \leftrightarrow,
\end{split}
\end{equation}
i.e., a product state composed of a two-photon bound state and a single photon which are centered at $a_2$ and $a_1$ respectively with $a_2<a_1$. Figure~\ref{fig:interactions} shows the evolution of the pulse delay for this state as it propagates through the ensemble, i.e. for different $N$. As in previous figures, a frame comoving with the pulse in the absence of interactions (i.e. $N=0$) is assumed. Here since $a_2<a_1$ and the Wigner delay is larger for single photons $\tau_1>\tau_2$, the bound-state photons catch up to and overtake the single photon. In this process the photons interact when the two parts overlap. The interaction causes a change in the Wigner delays which is seen as kinks in the lines in Figs.~\ref{fig:interactions}(a)-(b). The region of interaction is highlighted by the third-order correlations shown in Fig.~\ref{fig:interactions}(c). After the interaction has ended ($N \gtrsim 10$) the lines in Fig.~\ref{fig:interactions} (a) continue with the same slopes as before the interaction, signifying that there is still a two-photon bound state and an unbound photon. The collision between the bound state and the free photon is thus elastic and the bound state is stable against external influence.

\section{Outlook} \label{sec:outlook}

\begin{figure*}[!t]
\includegraphics[width=1.7\columnwidth]{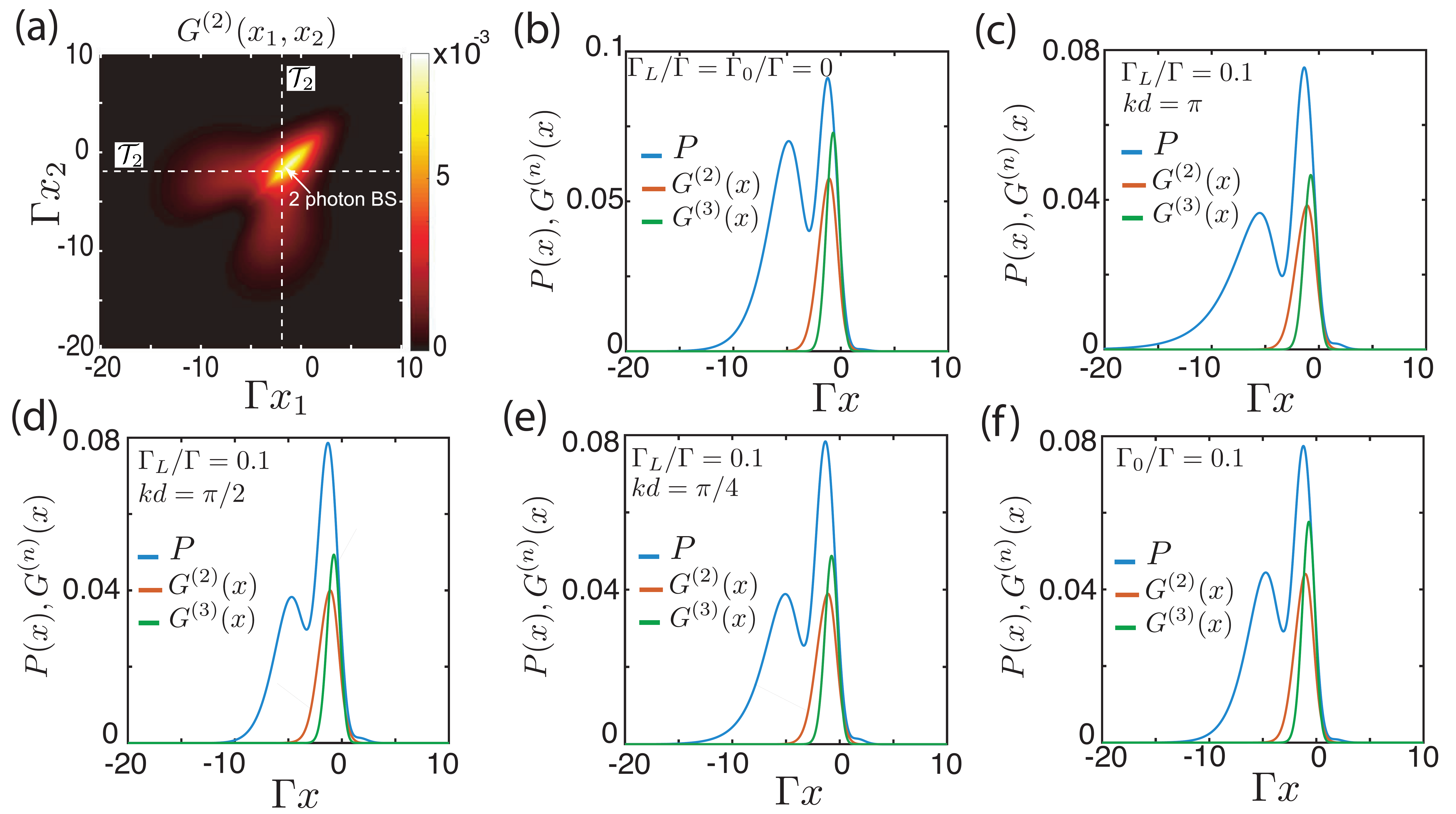}
\caption{\label{fig:N=2implement} Propagation of a coherent pulse with average photon number $\bar n=0.5$ through $N=2$ emitters. (a) Two-time correlation function $G^{(2)}(x_1, x_2)$ for pulse width $\Gamma \sigma= 3 \sqrt{2}$ for ideal chiral coupling computed using the three-photon theory.  Dashed lines show $N$ times the Wigner delay of the two-photon bound state $\tau_2$. (b) Power $P(x)$ and second- and third-order correlation functions $G^{(2)}(x)$ and $G^{(3)}(x)$ for pulsewidth $\Gamma \sigma = \sqrt{2}$ with ideal coupling.  Power and correlations when also coupling to a backward mode with rate $\Gamma_L = 0.1\Gamma$ where the emitters are separated by (c) $k d = \pi$, (d) $k d = \pi/2$, and (e) $k d = \pi/4$. (f) Power and correlation functions for unidirectional coupling, but with each emitter also coupled to a loss mode with $\Gamma_0 = 0.1 \Gamma$. Panels (b)-(f) are computed using a full numerical treatment.}
\end{figure*}

Our theoretical and numerical predictions have shown that many-body photon bound state propagation can be observed in chiral waveguide QED geometries with many emitters and photons. While these predictions are in the realm of quantum many-body physics, our work also predicts novel photon transport in the few-photon--few-emitter landscape. This is exemplified in Fig.~\ref{fig:N=2implement} where we consider coherent pulse propagation with average photon number $\bar n=0.5$ and $N=2$ emitters. Figures~\ref{fig:N=2implement}(a)-(b) show the the output power and correlation functions in the limit of ideal chiral coupling and no loss. Figure~\ref{fig:N=2implement}(a) shows the two-time correlation function $G^{(2)}(x_1, x_2)$ for an input pulse with width $\Gamma \sigma = 3 \sqrt{2}$. The width of this input pulse is chosen such that, in the two-excitation subspace, it projects on both the two-photon bound state subspace and the extended states with roughly equal probability. The distinct signatures of these two states can be seen: the bound state clearly propagates faster and is seen as an antinode on the diagonal marked by the intersection of the dashed lines. The spread out tails which propagate slower are the signature of the extended states. These are clearly not bunched in comparison to the bound state. Figure~\ref{fig:N=2implement}(b) shows the equal-time correlation function for a narrower pulse width $\Gamma \sigma= \sqrt{2}$. Here a narrower pulsewidth is chosen so it dominantly projects on the two-photon bound state. The hallmark of the photon--photon interactions is observed in the difference between the power $P(x)$ and the correlation functions $G^{(2)}(x)$, and $G^{(3)}(x)$. Clearly, the left-most peak, corresponding to the single-photon component, undergoes a larger time-delay than the bound states. The difference in time-delay between two- and three-photon bound states is also visible in the slight difference between the peak centers of $G^{(2)}(x)$, and $G^{(3)}(x)$. We note that it is also possible to observe a difference between $P(x)$ and $G^{(2)}(x)$ for a single quantum emitter.

We have also investigated the robustness of this effect when imperfections are introduced. Figures~\ref{fig:N=2implement}(c)-(e) consider additional coupling to the left-propagating mode at a rate $\Gamma_L=0.1\Gamma$ for different emitter spacings $k d$, where $k$ is the propagation wavenumber, and $d$ is the distance between each of the emitters. Note that unlike the fully chiral regime, when $\Gamma_L \neq 0$ the distance between emitters influences the dynamics of the system. Although the output field is slightly reduced in all three cases, the difference in the shape of the power and the correlation functions is preserved as is the difference in the peak positions. Finally, Fig.~\ref{fig:N=2implement}(f) considers ideal chiral coupling, but with each emitter coupling to a loss reservoir at a rate $\Gamma_0=0.1\Gamma$. Here, the single-photon component suffers the largest loss, while the bound states suffer a reduced loss. The introduction of the the loss does not spoil the difference in the shape of the power and the correlation functions.

The results here demonstrate that the propagation of few-photon bound states can be observed with as few as two quantum emitters chirally coupled to a reservoir and is robust to imperfections. This means that the phenomena shown here can be realized by several platforms currently under investigation. Optical quantum dots have demonstrated chiral coupling between a single emitter and a waveguide \cite{Sollner2015NNANO, Javadi2018NNANO}, while coupling between two diamond impurity centers and a photonic nanostructure has been achieved \cite{Sipahigil2016Science}. At microwave frequencies, circuit QED platforms can also achieve strong coupling between a superconducting qubit and a transmission line \cite{Hoi2012PRL} and a scheme for unidirecitonal coupling has been proposed \cite{Guimond2019arXiv}. Multiple qubits have also been coupled to a single mode or propagating modes \cite{vanLoo2013Science, Wen2019arXiv, Mirhosseini2019Nature,  Fitzpatrick2017PRX, Sundaresan2019PRX}. Finally, gasses of Rydberg atoms under the conditions of electromagnetically induced transparency also exhibit strong nonlinearities \cite{Firstenberg2013Nature} and possess bound eigenstates \cite{Maghrebi2015PRL}. Under certain limits they can also be mapped on to a nonlinear Schr\"{o}dinger equation with an attractive interaction \cite{Liang2017Science}. Such a Hamiltonian possesses bound eigenstates \cite{Thacker1981RMP, Lai1987PRAII} and therefore sufficiently long samples of Rydberg gasses can also potentially exhibit the bound state propagation shown here. Alternatively, ensembles can be engineered using Rydberg blockade to mimic chirally coupled emitters \cite{Stiesdal2018PRL}. The effects we predict here are thus widely applicable and can be observed in many different physical systems spanning vastly different energy scales.

Throughout this manuscript we have focused on understanding the fundamental physics of photonic bound state propagation. In addition to its fundamental interest, the dynamics of this system is highly interesting from an applied perspective. As a particular example we note that, for the output state shown in Fig.~\ref{fig:propagation}(a)-(d), if one selects a temporal window centered at $\Gamma x_0 = -N$, a state most likely containing either zero or two photons is produced. Launching the output on a beamsplitter and conditioning on the detection of a photon will thus produce a single-photon Fock state. For a small amplitude initial coherent state, the main contribution to this process will arise from the two-photon component of the incoming field. Since this is in a pure state, the outgoing single photon will also be in a pure state, although the temporal mode function will depend on the detection time. A detailed investigation of using photon bound-state propagation as a photon source is beyond the scope of this work. We can however estimate that for the parameters in Fig.~\ref{fig:propagation}(a)-(d), which are in no way optimized for the application, choosing a window with width $\Gamma x_w= 8$ centered on $\Gamma x_0 = -N$ leads to a probability of obtaining two photons  $P_2 = 0.061$ and an error probability of obtaining one or three photons $P_1+P_3 = 1.5\times 10^{-3}$. The efficiency and error of the source appear promising as they should be compared with the those of the input coherent state which are $P_2=0.076$ and $P_1+P_3 = 0.32$ respectively. 

\section{Conclusion}  \label{sec:conclusion}

Our results show that chiral wQED platforms provide a highly nonlinear medium suited for exploring nonlinear optics at the quantum level. From the most elementary light-matter interaction --- the interaction between photons and two-level systems --- emerges correlated photonic states. These bound states are truly distinct physical objects  with their own dispersion relation and are stable against external influence. Our work provides a clear recipe for how these features can potentially be observed in experiments. In the limit where the number of photons is high, the bound states approach the well-known soliton solution of SIT. Our full quantum description on the other hand covers the entire spectrum from few-photon quantum propagation to genuine quantum many-body (atom and photon) phenomena, and ultimately the quantum-to-classical transition. In particular, the analysis highlights  how the mean-field solution with weak quantum correlations breaks down through a region of maximal quantum correlations, until finally resulting in a state with weaker correlations. This work therefore paves the way for observing many-body quantum states of light in waveguide QED.

\section{Acknowledgements}

The authors would like to thank Mantas ~\v{C}epulkovskis, Peter Lodahl, Sebastian Hofferberth,  Hanna le Jeannic, Johannes Bjerlin, L. Henriet, and J. Douglas for fruitful discussions. S.M. and K.H. acknowledge funding from DFG through CRC 1227 DQ-mat, projects A05 and A06, and ``Nieders\"{a}chsisches Vorab'' through the ``Quantum-and Nano-Metrology (QUANOMET)''. G.C. and D.E.C. acknowledge support from ERC Starting Grant FOQAL, MINECO Severo Ochoa Grant SEV-2015-0522, CERCA Programme/Generalitat de Catalunya, Fundaci\'{o} Privada Cellex, Fundaci\'{o} Mir-Puig, Fundaci\'{o}n Ram\'{o}n Areces Project CODEC, European Quantum Flagship Project QIA, QuantumCAT (funded within the framework of the ERDF Operational Program of Catalonia), and Plan  Nacional Grant ALIQS, funded by MCIU, AEI, and FEDER. A.S.S. acknowledges support from the Danish National Research Foundation (Center of Excellence Hy-Q).

\bibliography{bigBib}

\onecolumngrid
\newpage

\pagebreak
\widetext
\begin{center}
\textbf{\large Supplemental Material: Dynamics of many-body photon bound states in chiral waveguide QED}
\end{center}
\setcounter{equation}{0}
\setcounter{figure}{0}
\setcounter{table}{0}
\setcounter{section}{0}

\makeatletter
\renewcommand{\theequation}{S\arabic{equation}}
\renewcommand{\thefigure}{S\arabic{figure}}
\renewcommand{\thesection}{S\arabic{section}}

\section{Self-induced Transparency}

Here we derive the fundamental soliton solution of self-induced transparency. While this calculation is covered in detail in \cite{Bullough1974OptoElectronics}, we include it here for completeness. The calculations in the main text set $v_g=1$. Here we include the group velocity in the calculations to keep track of the units of each expression. The Hamiltonian (\ref{eq:Hamiltonian}) becomes
\begin{equation}
\label{eq:SupplHamiltonian}
\hat{H} = -i v_g \int dx \, \hat{a}^\dagger (x) \frac{\partial}{\partial x} \hat{a}(x) + \sqrt{\Gamma v_g}\sum_{i=1}^N  \left[\hat{\sigma}_i^- \hat{a}^\dagger (x_i) + \hat{\sigma}_i^+ \hat{a} (x_i)\right].
\end{equation}
The two-level systems in SIT are modelled as a spin continuum. We do this via the mapping $\hat{\sigma}_i \rightarrow \hat{\sigma}(x)/\nu$, where $\hat{\sigma}_i$ is any of the Pauli operators for the $i$th TLS, and $\nu$ is the linear density of TLSs. This mapping produces the commutation relation 
\begin{equation}
\label{eq:ContinuumCommutators}
[\hat{\sigma}_- (x), \hat{\sigma}_+ (x')] = - \hat{\sigma}_z (x) \operatorname{\delta}{(x-x')}.
\end{equation}
We note that the continuum spin operators now have units of linear density. Within the continuum limit the Hamiltonian is 
\begin{equation}
\label{eq:ContinuumHamiltonian}
\hat{H} = -i v_g \int dx \, \hat{a}^\dagger (x) \frac{\partial}{\partial x} \hat{a}(x) + \sqrt{\Gamma v_g} \int dx  \left[\hat{\sigma}^-(x) \hat{a}^\dagger (x_i) + \hat{\sigma}^+(x) \hat{a} (x_i)\right].
\end{equation}
Obtaining the equations of motion for the operators and taking their expectation values within a mean-field approximation gives the SIT equations
\begin{equation}\label{eq:MFsuppl}
\begin{split}
\left[ \frac{\partial}{\partial t}  + \frac{\partial}{\partial x} \right] a(x,t) = - i \sqrt{\Gamma v_g}\sigma_-(x,t),\\
\frac{\partial}{\partial t} \sigma_-(x,t) = i \sqrt{\Gamma v_g} \sigma_z(x,t)a(x,t),\\
\frac{\partial}{\partial t} \sigma_z(x,t) = 4 \sqrt{\Gamma v_g} \operatorname{Im}{\left[a(x,t)\sigma_-^* (x,t) \right]}.
\end{split}
\end{equation}
Since we are considering a resonant pulse, $\sigma_-(x,t)$ is imaginary and we can thus set $\sigma_-(x,t) = i s(x,t)$. Examining the form of the last two equations of (\ref{eq:MFsuppl}), they can both be satisfied by defining
\begin{equation}
\begin{split}
s(x,t) &= -\frac \nu 2 \sin{\left[2 \chi(x,t)\right]},\\
\sigma_z(x,t) &= -\nu \cos{\left[2 \chi(x,t)\right]},
\end{split}
\end{equation}
where
\begin{equation}
\chi(x,t) = \sqrt{\Gamma v_g} \int_{-\infty}^t dt' a(x,t').
\end{equation}
Here $s(x,t)$ and $\sigma_z(x,t)$ are chosen such that in the limit of no field $\sigma_z(x,t) \rightarrow -\nu$, which is the desired physical limit. Noting that $a(x,t) = \partial \chi/\partial t/\sqrt{\Gamma v_g}$, the first equation in (\ref{eq:MFsuppl}) becomes
\begin{equation}
\label{eq:solitonGoverning1}
\left[ \frac{\partial^2}{\partial t^2}  + \frac{\partial^2}{\partial x \partial t} \right] \chi(x,t) = \frac{\Gamma v_g \nu}{2} \sin{\left[2 \chi(x,t) \right]}.
\end{equation}
Since we are looking for soliton solutions, we look for a functional form that propagates without changing shape. Such functions therefore only depend on a single variable $\xi = x - V' t$, where $V'$ is the propagation velocity of the soliton. One can then express (\ref{eq:solitonGoverning1}) as   
\begin{equation}
\label{eq:solitonGoverning2}
\frac{\partial^2}{\partial \xi^2} \chi(\xi) = \frac{\kappa^2}{2} \sin{\left[2 \chi(\xi) \right]},
\end{equation}
where
\begin{equation}
\kappa^2 = \frac{\Gamma v_g \nu}{V'(v_g - V')}.
\end{equation}
Equation (\ref{eq:solitonGoverning2}) has the solution $\chi(\xi) = \arcsin{\left[\tanh{(\kappa \xi)}\right]}$. This gives the solution
\begin{equation}
a(x,t) = \frac{V' \kappa}{\sqrt{\Gamma v_g}} \operatorname{sech}{\left[ \kappa (x-V' t) \right]}.
\end{equation}
This solution fulfills the criterion for the fundamental SIT soliton $2 \sqrt{\Gamma v_g}\int dx a(x,t)/V' = 2\pi$. An expression for the velocity can be obtained in terms of the average photon number $\bar n = v_g \int dx a(x,t)^2/V'$. This gives 
\begin{equation}
V' = \frac{\bar n^2 v_g \Gamma}{\bar n^2 \Gamma + 4 v_g \nu}.
\end{equation}
Substituting the velocity back into the expression for the field gives
\begin{equation}
a(x,t) = \frac{\bar n \sqrt{\Gamma}}{2 \sqrt{v_g}} \operatorname{sech}{\left[ \frac{\bar n \Gamma}{2} \left(\frac{x}{V'} -  t \right) \right]}.
\end{equation}
In the main text we consider a frame comoving with the pulse in the absence of atoms, i.e., at velocity $v_g$. Within this frame the pulse has a velocity in the backwards direction $V = v_g-V' = \frac{4 v_g^2 \nu}{\bar n^2 \Gamma} (1 + 4 v_g \nu/\bar n^2 \Gamma)^{-1}$. The delay per emitter can be obtained by noting that the velocity can be expressed in terms of the time it takes for the pulse to pass an atom $V' = \nu^{-1}/\left[(\nu v_g)^{-1} + \tau_{\bar n} \right]$. This gives an average delay per emitter of $\tau_{\bar n} = 4/(\bar n^2 \Gamma )$ which is  equivalent to the delay of the $n$-photon bound state.

\section{One-, Two- and Three-photon Transport using Scattering Eigenstates}

In this section we detail how we compute the output photon states when $N$ emitters are impinged upon by one, two, or three photons. We also combine these scattering calculations to consider scattering of a coherent state with up to three photons.

\subsection{Single-Photon Transport}

In the few-photon transport calculations we consider Gaussian input states. The normalized single-photon input state has the form
\begin{equation}
|\textrm{in} \rangle_1 = \int dx \, \hat{a}^{\dagger}(x) | 0 \rangle \frac{1}{{\sqrt{\sigma}} \pi^{\frac 1 4}} e^{i k_0 x}e^{-x^2/(2 \sigma^2)} 
\end{equation}
with its Fourier representation
\begin{equation}
|\textrm{in} \rangle_1 = \int dk \, \hat{a}^{\dagger}(k) | 0 \rangle \frac{\sqrt{\sigma}}{\pi^{1/4}} e^{-(k - k_0)^2 \sigma^2/2}. 
\end{equation}
The creation operators $\hat{a}^{\dagger}(k)$ are eigenstates of the single-photon scattering operator. After $N$ scattering events, we have $\hat{a}^{\dagger}(k) \rightarrow t_k^N \hat{a}^{\dagger}(k)$, with the transmission coefficient
\begin{equation}
t_k = \frac{k - i \Gamma/2}{ k + i \Gamma/2}.
\end{equation}
Throughout we consider an ideal 1D reservoir without losses, and with $v_g=1$. The scattered state is thus
\begin{equation}
\label{eq:singlePhotOut}
|\textrm{out} \rangle_1 = \sqrt{\frac{\sigma}{2 \pi}} \frac{1}{\pi^{1/4}} \int dx \, \hat{a}^{\dagger}(x) | 0 \rangle \int dk \, t_k^N e^{-(k-k_0)^2 \sigma^2/2} e^{i k x}. 
\end{equation}
This is computed numerically. Most of the integrals that appear in the few-photon transport calculations are Fourier integrals, which can be efficiently computed using Fast Fourier Transforms.

\subsection{Two-Photon Transport}

The two-photon scattering matrix has two types of eigenstates, extended states $| W_{E, \Delta} \rangle$ and bound states $| B_E \rangle$ \cite{Shen2007PRA}. The scattering matrix has the form
\begin{equation}
\hat{S}_{22}^N = \frac{1}{2} \int dE d\Delta \, t_{\frac E 2 + \Delta}^N t_{\frac E 2 - \Delta}^N | W_{E, \Delta} \rangle_2 \langle W_{E, \Delta} | + \int dE \, t_{E,2}^N | B_{E} \rangle_2 \langle B_{E} |,
\end{equation}
where all integrals range from $-\infty$ to $\infty$. The two-photon scattering eigenstates can be obtained from Bethe's ansatz in the main text (\ref{eq:MBScattEig}). In a position-space representation they are
\begin{equation}
\begin{split}
|W_{E,\Delta} \rangle_2 &= \frac{1}{\sqrt{2}}\int dx_1 dx_2  \hat{a}^\dagger({x_1}) \hat{a}^\dagger ({x_2})| 0 \rangle \, W_{E,\Delta} (x_c,x)\\
|B_{E} \rangle_2 &= \frac{1}{\sqrt{2}}\int dx_1 dx_2  \hat{a}^\dagger({x_1}) \hat{a}^\dagger ({x_2})| 0 \rangle \, B_{E} (x_c,x).
\end{split}
\end{equation}
Here,
\begin{equation}
\begin{split}
\label{eq:twoPhot}
W_{E,\Delta} (x_c,x) &= \frac{1}{\sqrt{4 \Delta^2 + \Gamma^2}}\frac{\sqrt{2}}{2 \pi}e^{i E x_c}\left[2 \Delta \cos{(\Delta x)} - \Gamma \operatorname{sgn}{(x)}\sin{(\Delta x)} \right],\\
B_{E} (x_c,x) &= \sqrt{\frac{\Gamma}{4\pi}}\,e^{i E x_c}\,e^{-\frac \Gamma 2 |x|},\\
\end{split}
\end{equation}
and
\begin{equation}
t_{E,2} = \frac{E - 2i \Gamma}{E + 2i \Gamma},
\end{equation}
and $x_c=\frac{x_1+x_2}{2}$, $x=x_1-x_2$, $E=k+p$ is a two-photon detuning, and $\Delta=\frac{k-p}{2}$ is a difference in photon energies, where $k$ and $p$ are photon detunings.

The input state for two photons is
\begin{equation}
\begin{split}
|\textrm{in} \rangle_2 &= \frac{1}{\sqrt{2}}\int dx_1 dx_2 \, \hat{a}^{\dagger}(x_1) \hat{a}^{\dagger}(x_2) | 0 \rangle \frac{1}{\sigma \sqrt{\pi}} e^{ - x_1^2/2 \sigma^2} e^{ - x_2^2/2 \sigma^2} e^{i k_0 (x_1 + x_2)}\\
&= \frac{1}{\sqrt{2}}\int dx_1 dx_2 \, \hat{a}^{\dagger}(x_1) \hat{a}^{\dagger}(x_2) | 0 \rangle \frac{1}{\sigma \sqrt{\pi}} e^{ - x_c^2/ \sigma^2} e^{ - x^2/4 \sigma^2} e^{2 i k_0 x_c}.
\end{split}
\end{equation}
Projecting this on the two-photon states gives
\begin{equation}
{}_2\langle B_E | \textrm{in} \rangle_2 =  \sqrt{\Gamma} \sigma e^{-(E-2k_0)^2 \sigma^2/4} e^{\Gamma^2 \sigma^2/4} \operatorname{Erfc}{\left( \frac{\Gamma \sigma}{2}\right)}
\end{equation}
and
\begin{equation}
{}_2\langle W_{E,\Delta} | \textrm{in} \rangle_2 =  \frac{4 \sigma}{\pi \sqrt{2}} \frac{1}{\sqrt{4 \Delta^2 + \Gamma^2}} e^{ - (E - 2 k_0)^2 \sigma^2 / 4} \left[ \sqrt{\pi} \Delta e^{- \Delta^2 \sigma^2} - \Gamma D_F(\Delta \sigma) \right],
\end{equation}
where $\textrm{Erfc} = 1- \textrm{Erf}$, where $\textrm{Erf}$ is the error function, and $D_F$ is Dawson's Function.

The output state is then written in terms of two integrals given by the bound-state and extended-state contributions
\begin{equation}
|\textrm{out} \rangle_2 = \frac{1}{\sqrt{2}} \int dx_1 dx_2 \, \hat{a}^\dagger (x_1) \hat{a}^\dagger (x_2) | 0 \rangle \left[ \psi_B(x_c, x) + \psi_W(x_c, x) \right],
\end{equation}
with
\begin{equation}
\psi_B(x_c, x) = \sqrt{\Gamma}  \sigma \sqrt{\frac{\Gamma}{4 \pi}}e^{\Gamma^2 \sigma^2/4}  \operatorname{Erfc}{\left( \frac{\Gamma \sigma}{2}\right)}  e^{-\frac{\Gamma}{2}|x|} \int dE \, t_{E,2}^N e^{- (E- 2k_0)^2 \sigma^2/4}e^{i E x_c},
\end{equation}
and 
\begin{equation}
\begin{split}
\psi_W(x_c, x) &= \frac{\sigma}{\pi^2} \int dE d\Delta \, t_{\frac E 2 + \Delta}^N t_{\frac E 2 - \Delta}^N  e^{-(E - 2k_0)^2 \sigma^2/4} e^{i E x_c} \frac{1}{4 \Delta^2 + \Gamma^2} \left[ \sqrt{\pi} \Delta e^{ - \Delta^2 \sigma^2} - \Gamma D_F(\Delta \sigma)\right]\\
& \times \left[2 \Delta \cos{(\Delta x)} - \Gamma \operatorname{sgn}{{(x)}} \sin{(\Delta x)} \right].
\end{split}
\end{equation}
The integrals in these two functions are computed numerically.

\subsection{Three-photon transport}

Using the same rules rules we construct the three-photon scattering matrix for $N$ emitters    
\begin{equation}
\label{eq:threePhotScattMat}
\hat{S}_{33}^N = \frac{1}{3!} \int d^3 \mathbf k \, t_{\mathbf k}^N \, |W_{\mathbf k} \rangle_3 \langle W_{\mathbf k} | + \frac{1}{2} \int dK dk \, t_{k}^N t_{K,2}^N |H_{K, k} \rangle_3 \langle H_{K, k} | + \int dE \, t_{E,3}^N \, |B_{E} \rangle_3 \langle B_{E} |,
\end{equation}
where
\begin{equation}
t_{E,3} = t_{E/3 - i \Gamma} t_{E/3} t_{E/3 + i \Gamma} = \frac{E - 9 i \Gamma/2}{E + 9 i \Gamma/2}.
\end{equation}
The eigenstates are
\begin{equation}
\begin{split}
|W_{\mathbf k} \rangle_3 &= \frac{1}{\sqrt{3!}}\int dx_1 dx_2 dx_3  \hat{a}^\dagger({x_1}) \hat{a}^\dagger ({x_2})  \hat{a}^\dagger ({x_3})| 0 \rangle \, W_{\mathbf k, 3} (x_1, x_2, x_3)\\
|H_{K, k} \rangle_3 &= \frac{1}{\sqrt{3!}}\int dx_1 dx_2 dx_3  \hat{a}^\dagger({x_1}) \hat{a}^\dagger ({x_2})  \hat{a}^\dagger ({x_3})| 0 \rangle \, H_{K, k, 3} (x_1, x_2, x_3)\\
|B_{E} \rangle_3 &= \frac{1}{\sqrt{3!}}\int dx_1 dx_2 dx_3  \hat{a}^\dagger({x_1}) \hat{a}^\dagger ({x_2})  \hat{a}^\dagger ({x_3})| 0 \rangle \, B_{E, 3} (x_1, x_2, x_3)\\
\end{split}
\end{equation}
with the normalized real-space representations,
\begin{equation}
\begin{split}
\label{eq:threePhotEigenstates}
W_{\mathbf k, 3} (x_1, x_2, x_3) &= \frac{1}{\sqrt{3! (2\pi)^3 \prod_{m<n}^{3} \left[ (k_m - k_n)^2 + \Gamma^2 \right]}} \left\{\prod_{m<n}^3 \left[k_m - k_n - i \Gamma \operatorname{sgn}{(x_{n} - x_{m})} \right] \prod_{j=1}^3 e^{i k_j x_{j}}\right\} + \leftrightarrow, \\
H_{K, k, 3} (x_1, x_2, x_3) &= \frac{- i \Gamma }{2\pi \sqrt{3 \Gamma \left[(\frac K 2 - k)^2 + \frac{\Gamma^2}{4} \right] \left[(\frac K 2 - k)^2 + \frac{9\Gamma^2}{4} \right]}} \theta(x_{2} - x_{1}) \left[\frac K 2 - k - \frac{i \Gamma }{2} - i \Gamma \operatorname{sgn}{(x_{3} - x_{1})} \right]\\
& \times \left[\frac K 2 - k + \frac{i \Gamma }{2} - i \Gamma \operatorname{sgn}{(x_{3} - x_{2})} \right] e^{i \frac K 2 (x_{1} + x_{2})} e^{ - \frac \Gamma 2 (x_{2} - x_{1})} e^{i k x_{3}} +  \leftrightarrow,\\
B_{E, 3} (x_1, x_2, x_3) &= \frac{\Gamma}{\sqrt{3 \pi}} e^{i E (x_1+x_2+x_3)/3} e^{- \Gamma/2 (|x_1 - x_2| + |x_1 - x_3| + |x_2 - x_3|)}.\\
\end{split}
\end{equation}

The three-photon Gaussian input state is
\begin{equation}
\label{eq:threePhotInputGauss}
| \textrm{in} \rangle_3 = \frac{1}{\sqrt{3!}} \int dx_1 dx_2 dx_3 \hat{a}^\dagger (x_1) \hat{a}^\dagger (x_2) \hat{a}^\dagger (x_3) | 0 \rangle \frac{1}{\pi^{3/4}\sigma^{3/2}} e^{i k_0 (x_1 + x_2 + x_3)} e^{- x_1^2/(2 \sigma^2)} e^{- x_2^2/(2 \sigma^2)} e^{- x_3^2/(2 \sigma^2)}.
\end{equation}
Both the projection of the input state on the scattering eigenstates and the subsequent integrals in (\ref{eq:threePhotScattMat}) are performed numerically. 

\begin{figure}[!t]
\includegraphics[width=0.5\columnwidth]{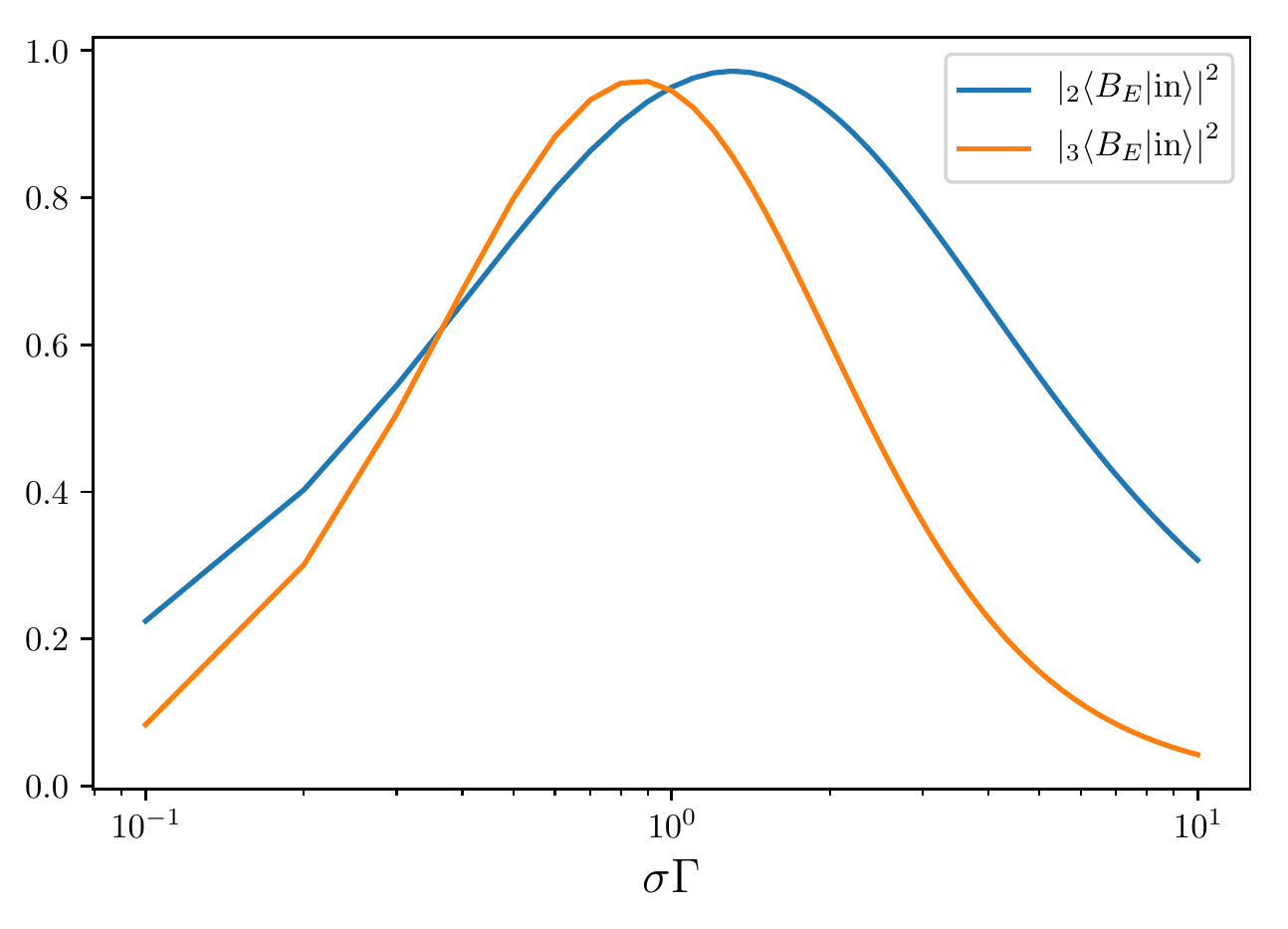}
\caption{\label{fig:projGaussOnBound} Overlap of a two- and three-photon Gaussian states with two- and three-photon bound states versus the width of the Gaussian state $\sigma \Gamma$. }
\end{figure}

The total contribution of the bound state varies with the input pulse width. Figure~\ref{fig:projGaussOnBound} shows the overlap between the three-photon bound state and an input Gaussian versus pulsewidth. We note that the maximum overlap for the three-photon bound state occurs for shorter pulses than the two-photon bound state.

\subsection{Output states and observables}

We now have one-, two-, and three-photon output states which we express as
\begin{equation}
\begin{split}
|\textrm{out} \rangle_1 &= \int dx_1 \, \hat{a}^\dagger (x_1) | 0 \rangle \psi_1(x_1),\\
|\textrm{out} \rangle_2 &= \frac{1}{\sqrt{2}}\int dx_1 dx_2 \, \hat{a}^\dagger (x_1) \hat{a}^\dagger (x_2) | 0 \rangle \psi_2(x_1, x_2),\\
|\textrm{out} \rangle_3 &= \frac{1}{\sqrt{3!}}\int dx_1 dx_2 dx_3 \, \hat{a}^\dagger (x_1) \hat{a}^\dagger (x_2) \hat{a}^\dagger (x_3) | 0 \rangle \psi_3(x_1, x_2, x_3).
\end{split}
\end{equation}
These are also used to compute the output for coherent states
\begin{equation}
\begin{split}
| \alpha_{\rm out} \rangle = e^{- |\alpha|^2 / 2} \left[| 0 \rangle + \alpha | \textrm{out} \rangle_1 +  \frac{\alpha^2}{\sqrt{2}} | \textrm{out} \rangle_2  + \frac{\alpha^3}{\sqrt{3!}} | \textrm{out} \rangle_3  + \ldots \right]
\end{split}
\end{equation}

We are interested in computing the normally ordered expectation values and correlations $\langle \hat{a}^\dagger(x) \hat{a}(x) \rangle$, $G^{(2)}(x_c, x) = \langle \hat{a}^\dagger(x_c) \hat{a}^\dagger(x_c + x) \hat{a}(x_c+x) \hat{a}(x_c) \rangle$, and $G^{(3)}(x_c, x, y) = \langle \hat{a}^\dagger(x_c) \hat{a}^\dagger(x_c + x) \hat{a}^\dagger(x_c + y) \hat{a}(x_c+y) \hat{a}(x_c+x) \hat{a}(x_c) \rangle$. For a coherent state with up to three photons, these give the expressions
\begin{equation}
\begin{split}
\langle \hat{a}^\dagger(x) \hat{a}(x) \rangle &= |\alpha|^2 |\psi_1 (x)|^2 + |\alpha|^4 \left[\int dx_1 |\psi_2 (x, x_1)|^2 - |\psi_1(x)|^2 \right]\\
& + |\alpha|^6 \left[ \frac 1 2 \int dx_1 dx_2 |\psi_3 (x, x_1, x_2)|^2  - \int dx_1 |\psi_2(x, x_1)|^2 + \frac 1 2 |\psi_1(x)|^2 \right],\\
G^{(2)}(x_c, x) &= |\alpha|^4 |\psi_2 (x_c, x_c+x)|^2 + |\alpha|^6 \left[  \int dx_1 |\psi_3 (x_c, x_c+x, x_1)|^2 - |\psi_2 (x_c, x_c+x)|^2\right],\\
G^{(3)}(x_c, x, y) &= |\alpha|^6 |\psi_3 (x_c, x_c +x, x_c +y)|^2.
\end{split}
\end{equation}

\section{Separability of three-photon hybrid states}

The behaviour of the hybrid state in Fig.~\ref{fig:propagation}(c) of the main text is significant. Its shape indicates that the bound two-photon part of this state propagates with the same effective velocity as the actual two-photon bound state, while the extended photon in the hybrid state evolves like a single photon. 

In this section we show that, for spectrally narrow input states, the evolution of the Hybrid state follows that of an independent two-photon bound state and an independent single photon. We consider the input state $|\textrm{in} \rangle_3$ given in (\ref{eq:threePhotInputGauss}) that is on resonance $k_0=0$. The hybrid state has the form given in (\ref{eq:threePhotEigenstates}). The contribution of the hybrid state to the full three-photon output state is
\begin{equation}
\begin{split}
\label{eq:hybridPartWvFunc}
| \textrm{out} \rangle_{\rm hybrid} &= \frac 1 2 \int dK dk \, t_{K,2}^N t_k^N| H_{K,k} \rangle_3 \langle H_{K,k} | \textrm{in} \rangle\\
&= \frac{1}{\sqrt{3!}}\int dx_1 dx_2 dx_3 \psi_{\rm hybrid} (x_1,x_2,x_3) \hat{a}^\dagger (x_1) \hat{a}^\dagger (x_2) \hat{a}^\dagger (x_3) | 0 \rangle
\end{split}
\end{equation}

We now focus on the inner product $\langle H_{K,k} | \textrm{in} \rangle$. Noting that the input state is separable and therefore already fully symmetric we can relabel the $3!$ terms in the Hybrid state to make them identical. We end up with
\begin{equation}
\begin{split}
{}_3\langle \textrm{in}  | H_{K,k} \rangle_3 &= \frac{- i \Gamma }{2\pi \sqrt{3 \Gamma \left[(\frac K 2 - k)^2 + \frac{\Gamma^2}{4} \right] \left[(\frac K 2 - k)^2 + \frac{9\Gamma^2}{4} \right]}} \frac{3!}{\pi^{3/4}\sigma^{3/2}} \int dx_1 dx_2 dx_3 \,  e^{-(x_1^2+x_2^2 + x_3^2)/(2\sigma^2)}  \theta(x_{2} - x_{1})\\
& \left[\frac K 2 - k - \frac{i \Gamma }{2} - i \Gamma \operatorname{sgn}{(x_{3} - x_{1})} \right] \left[\frac K 2 - k + \frac{i \Gamma }{2} - i \Gamma \operatorname{sgn}{(x_{3} - x_{2})} \right] e^{i \frac K 2 (x_{1} + x_{2})} e^{ - \frac \Gamma 2 (x_{2} - x_{1})} e^{i k x_{3}}.
\end{split}
\end{equation}
Using the two-body coordinates $x_c = x_1+x_2/2$ and $x=x_1-x_2$ allows writing the integral as
\begin{equation}
\begin{split}
\label{eq:bigHybridIntegral}
I(K,k) &= \int dx_c dx \, dx_3 \,  e^{-(2 x_c^2 + x^2/2 + x_3^2)/(2\sigma^2)}  \theta(-x)
\left[\frac K 2 - k - \frac{i \Gamma }{2} - i \Gamma \operatorname{sgn}{\left(x_{3} - x_c - \frac x 2 \right)} \right] \\
&\left[\frac K 2 - k + \frac{i \Gamma }{2} - i \Gamma \operatorname{sgn}{\left(x_{3} - x_c + \frac x 2 \right)} \right] e^{i K x_c} e^{ \frac \Gamma 2 x} e^{i k x_{3}}\\
=& \int dx_c dx_3 e^{-(2 x_c^2 + x_3^2)/(2\sigma^2)} e^{i K x_c} e^{i k x_{3}} \int  dx \, e^{ \frac \Gamma 2 x} e^{-x^2/(4 \sigma^2)} \,    \theta(-x) \left[\frac K 2 - k - \frac{i \Gamma }{2} - i \Gamma \operatorname{sgn}{\left(x_{3} - x_c - \frac x 2 \right)} \right] \\
&\left[\frac K 2 - k + \frac{i \Gamma }{2} - i \Gamma \operatorname{sgn}{\left(x_{3} - x_c + \frac x 2\right)} \right].
\end{split}
\end{equation}
Unfortunately all the integrals cannot be computed exactly. We now perform the integrals over the $x$ coordinate by considering $\Gamma \sigma \gg 1$. The exponentially decaying functions decay with $\Gamma$, while the Gaussian functions have a width $\sim 1/\sigma$. When taking $\Gamma \sigma \gg 1$, we can make the approximation $e^{\Gamma/2 x}e^{-x^2/(2\sigma^2)} \sim e^{\Gamma/2 x}$ for $x<0$. We can then write the integral over $x$ as
\begin{equation}
\begin{split}
&\int  dx \, e^{ \frac \Gamma 2 x} \, \theta(-x) \left[\left(\frac K 2 - k \right)^2 + \frac{\Gamma^2}{4} - \frac{\Gamma^2}{2} \operatorname{sgn}{\left(\frac x 2 - a \right)} + i \Gamma \left(\frac K 2 - k\right) \operatorname{sgn}{\left(\frac x 2 + a \right)} \right.\\
& \left. -  i \Gamma \left(\frac K 2 - k \right) \operatorname{sgn}{\left(\frac x 2 - a \right)} - \frac{\Gamma^2}{2} \operatorname{sgn}{\left(\frac x 2 + a \right)}  + \Gamma^2 \operatorname{sgn}{\left(\frac x 2 - a \right)} \operatorname{sgn}{\left(\frac x 2 + a \right)} \right],
\end{split}
\end{equation}
where $a = x_3 - x_c$. We then compute the integrals
\begin{equation}
\begin{split}
I_0 &= \int dx \, \theta(-x) e^{\Gamma / 2 x} = \frac{2}{\Gamma}\\
I_1 &= \int dx \, \theta(-x) e^{\Gamma / 2 x} \operatorname{sgn}{\left(\frac x 2 + a \right)} = \frac{2}{\Gamma} \left[(1-2 e^{-a \Gamma})\theta(a) - \theta(-a) \right]\\
I_2 &= \int dx \, \theta(-x) e^{\Gamma / 2 x} \operatorname{sgn}{\left(\frac x 2 - a \right)} = \frac{2}{\Gamma} \left[(1-2 e^{a \Gamma})\theta(-a) - \theta(a) \right]\\
I_3 &= \int dx \, \theta(-x) e^{\Gamma / 2 x} \operatorname{sgn}{\left(\frac x 2 - a \right)} \operatorname{sgn}{\left(\frac x 2 + a \right)} =  \frac 2 \Gamma \left(2 e^{- \Gamma |a|}-1 \right).
\end{split}
\end{equation}
The integral in (\ref{eq:bigHybridIntegral}) then becomes
\begin{equation}
\begin{split}
I(K,k) = &\frac 2 \Gamma \int dx_c dx_3 e^{-(2 x_c^2 + x_3^2)/(2\sigma^2)} e^{i K x_c} e^{i k x_{3}} \left\{ \left(\frac{K}{2}-k \right)^2 + \frac{\Gamma^2}{4} + \Gamma^2 e^{-\Gamma |a|} \right.\\
& \left. - 2 i \Gamma \left(\frac{K}{2} -k \right) \operatorname{sgn}{(a)} (1- e^{-\Gamma |a|}) + \Gamma^2 (2 e^{-\Gamma |a|} - 1)\right\},
\end{split}
\end{equation}
where we have used $I_1 + I_2 = -4/\Gamma e^{-\Gamma |a|}$ and $I_1 - I_2 = 4/ \Gamma \operatorname{sgn}{(a)} (1- e^{-\Gamma |a|})$. Unfortunately we are unable to compute $I(K,k)$ exactly. Instead we compute the projection of this function on the $K$-axis, i.e. $\tilde{I}(K) = \int dk I(K,k)$. We show that this function is much narrower than the linewidth. Consequently, conservation of energy $E=K+k$ will imply that its projection on the $k$-axis $\bar{I}(k) = \int dK I(K,k)$ is also narrower than the linewidth. After computing the integrals and some manipulation, in the limit $\Gamma \sigma \gg 1$, we obtain
\begin{equation}
\begin{split}
I(K) = \pi \sqrt{\pi}e^{- K^2 \sigma^2 /4}\left(\frac{K^2 \sigma^2}{\Gamma \sigma} -3 \Gamma \sigma + \frac{4}{\Gamma \sigma} \right) + \frac{8 \pi \Gamma^2}{K^2 + \Gamma^2} + 4 \pi K \left[\frac{2K}{K^2 + \Gamma^2} - \sqrt{\pi} \sigma e^{-K^2 \sigma^2/4} \operatorname{Erfi}{\left(\frac{K\sigma}{2} \right)} \right],
\end{split}
\end{equation}
where $\operatorname{Erfi}$ is the imaginary error function (which is actually real-valued). In the limit $K \rightarrow 0$ and $\Gamma \sigma \gg 1$ the function is
\begin{equation}
\begin{split}
\label{eq:IKapprox}
I(K) = \pi \sqrt{\pi} \left(-3 \Gamma \sigma +\frac{4}{\Gamma \sigma} \right) e^{- K^2 \sigma^2 /4} + O(K^2 \sigma^2).
\end{split}
\end{equation}
On the other hand, in the limit $K \sigma \gg 1$ and $\Gamma \sigma \gg 1$, using the large argument approximation $e^{-x^2/4}\operatorname{Erfi}{(x/2)}  = 2/(\sqrt{\pi} x) + O(1/x^3)$ we have
\begin{equation}
\begin{split}
I(K) = \pi \sqrt{\pi}e^{- K^2 \sigma^2 /4}\left(\frac{K^2 \sigma^2}{\Gamma \sigma} -3 \Gamma \sigma + \frac{4}{\Gamma \sigma}\right) + O\left(\frac{1}{K^2 \sigma^2}\right).
\end{split}
\end{equation}
The Gaussian part becomes negligible here and the function therefore scales as $1/K^2 \sigma^2$ which is much less than one when $K \rightarrow \Gamma$. Within the linewidth of the emitter, the function is largest when  $K \sigma  \ll 1$, where it is well-approximated by a Gaussian function. Since the tails of this function scales as $1/K^2 \sigma^2$, they only serve to slightly broaden the Gaussian function. This implies that (\ref{eq:IKapprox}) is a reasonable approximation for the function.

This means that the Gaussian part of the function is dominant over the other parts. This part of the function $I(K,k)$ comes from the integrals $I_0$ and $I_3$ and can actually be computed analytically when $\Gamma \sigma \gg 1$. It is given by 
\begin{equation}
\begin{split}
I(K,k) &\sim \frac 2 \Gamma \left[\left(\frac{K}{2} - k\right)^2 - \frac{3\Gamma^2}{4} \right]\sqrt{2}\pi \sigma^2 e^{-K^2 \sigma^2/4} e^{-k^2 \sigma^2/2}\\
& \sim \frac{3\Gamma}{2} \sqrt{2}\pi \sigma^2 e^{-K^2 \sigma^2/4} e^{-k^2 \sigma^2/2}.
\end{split}
\end{equation}
This then gives
\begin{equation}
\langle \textrm{in}  | H_{K,k} \rangle \sim \frac{- i \Gamma }{2\pi \sqrt{3 \Gamma \left[(\frac K 2 - k)^2 + \frac{\Gamma^2}{4} \right] \left[(\frac K 2 - k)^2 + \frac{9\Gamma^2}{4} \right]}} \frac{3!}{\pi^{3/4}\sigma^{3/2}} \frac{3\Gamma}{2} \sqrt{2}\pi \sigma^2 e^{-K^2 \sigma^2/4} e^{-k^2 \sigma^2/2}.
\end{equation}
This projection has two very useful features. The Gaussian part is separable in $K$ and $k$ and this part is highly localized about the origin. In fact the function multiplying the Gaussians can be approximated by a constant in $K$ and $k$ when $\Gamma \sigma \gg 1$.

Substituting the projection $\langle  H_{K,k} | \textrm{in} \rangle$ into the (\ref{eq:hybridPartWvFunc}) gives the hybrid wavefunction
\begin{equation}
\begin{split}
&\psi_{\rm hybrid}(x_1,x_2,x_3) = \frac 1 2 \int dK dk \, t_{K,2}^N t_k^N \frac{ \Gamma^2 }{12 \pi^2 \Gamma \left[(\frac K 2 - k)^2 + \frac{\Gamma^2}{4} \right] \left[(\frac K 2 - k)^2 + \frac{9\Gamma^2}{4} \right]} \frac{3!}{\pi^{3/4}\sigma^{3/2}} \frac{3\Gamma}{2} \sqrt{2}\pi \sigma^2 e^{-K^2 \sigma^2/4} e^{-k^2 \sigma^2/2} \\
& \left\{ \theta(x_{2} - x_{1}) \left[\frac K 2 - k - \frac{i \Gamma }{2} - i \Gamma \operatorname{sgn}{(x_{3} - x_{1})} \right] \left[\frac K 2 - k + \frac{i \Gamma }{2} - i \Gamma \operatorname{sgn}{(x_{3} - x_{2})} \right] e^{i \frac K 2 (x_{1} + x_{2})} e^{ - \frac \Gamma 2 (x_{2} - x_{1})} e^{i k x_{3}} +  \leftrightarrow \right\}.\\
&\sim \frac{ 4\Gamma^2 \sigma^2 \sqrt{2} \pi }{\Gamma^2} \frac{1}{\pi^{3/4}\sigma^{3/2}} \frac 1 2 \int dK dk \, t_{K,2}^N t_k^N   e^{-K^2 \sigma^2/4} e^{-k^2 \sigma^2/2} \\
& \left\{ \theta(x_{2} - x_{1})   \left[  \frac{ 1 }{2} + \operatorname{sgn}{(x_{3} - x_{1})} \right]  \left[\frac{ 1 }{2} -   \operatorname{sgn}{(x_{3} - x_{2})} \right] e^{i \frac K 2 (x_{1} + x_{2})} e^{ - \frac \Gamma 2 (x_{2} - x_{1})} e^{i k x_{3}} +  \leftrightarrow \right\}.
\end{split}
\end{equation}
The integrals over $K$ and $k$ are separable and can be computed individually. They are simply inverse Fourier transforms of the Gaussian functions multiplied by their transmission coefficients. The two-photon bound part of the hybrid state therefore evolves according to the two-photon bound state transmission coefficient, while the single photon part evolves according to the single-photon transmission coefficient. One can therefore interpret the hybrid eigenstates as a two-photon bound state and a single photon which are weakly interacting. The effect of interactions then becomes negligible in the long pulse limit, since the two components are unlikely to be at the same spatial positions.

\section{Electric field of bound states and relation to self-induced transparency}

In this section we compute the electric field of a linear combination of bound states. In particular, we show that in the classical limit, where the sum is composed of a sum of many-body bound states, the field resembles that of a self-induced transparency soliton.

We begin by considering an arbitrary state composed of a sum of bound states
\begin{equation}
| \psi \rangle = \sum_n \int dE  c_n(E) | B^E \rangle_n
\end{equation}
The electric field is then proportional to
\begin{equation}
\langle \hat{a}(x) \rangle = \langle \psi | \hat{a}(x) | \psi \rangle = \sum_n \int dE dE'  c^*_{n-1}(E') c_n(E) {}_{n-1}\langle B^E | \hat{a}(x) | B^E \rangle_n.
\end{equation}
The bulk of this calculation involves computing the matrix element in the integrand. Using the expression for the $n$-photon bound state in (\ref{eq:MBBound}) of the main text, the matrix element is
\begin{equation}
\begin{split}
{}_{n-1}\langle B^E | \hat{a}(x) | B^E \rangle_n =& \frac{\sqrt{n}}{2\pi} \sqrt{\frac{\Gamma^{n-2}(n-2)!}{n-1}}\sqrt{\frac{\Gamma^{n-1}(n-1)!}{n}} \int dx_1 dx_2 \ldots dx_{n-1} e^{i \left(\frac{E}{n} - \frac{E'}{n-1} \right) (x_1+x_2+\ldots x_{n-1})}\\
&\times e^{-\Gamma \sum_{i<j}^{n-1}|x_i - x_j|}e^{i E x/n}e^{-\frac{\Gamma}{2} \sum_i |x_i - x|}\\
=& \frac{\sqrt{\Gamma}}{2\pi} \Gamma^{n-2} (n-2)! \, e^{i E x/n} \int dx_1 dx_2 \ldots dx_{n-1} e^{i K_n(x_1 +x_2 +\ldots x_{n-1})/n} e^{-\Gamma \sum_{i<j}^{n-1}|x_i - x_j|}e^{-\frac{\Gamma}{2} \sum_i |x_i - x|},
\end{split}
\end{equation}
where $K_n = E - \frac{n E'}{n-1}$.

The bound-state wave functions are highly symmetric, which can be used to greatly restrict the domain of integration. This symmetry will be exploited to reduce the exponentials with negative absolute value arguments to exponentials without absolute values, but with arguments that are always negative. Noting that the integral is invariant under permutation of any $x_i \leftrightarrow x_j$, we can restrict the domain of integration to $x_1>x_2>\ldots x_{n-1}$ and correspondingly multiply by $(n-1)!$. This gives  
\begin{equation}
\begin{split}
{}_{n-1}\langle B^E | \hat{a}(x) | B^E \rangle_n &= \frac{\sqrt{\Gamma}}{2\pi} \Gamma^{n-2} (n-2)! (n-1)! \, e^{i E x/n} \int_> dx_1 dx_2 \ldots dx_{n-1} e^{i K_n(x_1 +x_2 +\ldots x_{n-1})/n}e^{-\frac{\Gamma}{2} \sum_i |x_i - x|}\\
& \times e^{-\Gamma(x_1 - x_2)} e^{-\Gamma(x_1 - x_3)} e^{-\Gamma(x_1 - x_4)} \times \ldots \times e^{-\Gamma(x_1 - x_{n-1})}\\
& \hspace{16mm} \times e^{-\Gamma(x_2 - x_3)} e^{-\Gamma(x_2 - x_4)} \times \ldots \times e^{-\Gamma(x_2 - x_{n-1})}\\
& \hspace{16mm} \hspace{16mm} \vdots \hspace{40mm} \vdots\\
& \hspace{34mm} \times e^{-\Gamma(x_{n-3}- x_{n-2})} e^{-\Gamma(x_{n-3}- x_{n-1})}\\
& \hspace{57mm} \times  e^{-\Gamma(x_{n-2}- x_{n-1})},
\end{split}
\end{equation}
where the subscript $>$ of the integral indicates the domain $x_1>x_2> \ldots > x_{n-1}$.  All these exponents can be combined to give
\begin{equation}
e^{-\Gamma(n-2)x_1} e^{-\Gamma(n-4)x_2} e^{-\Gamma(n-6)x_3} \ldots e^{-\Gamma\left[n-2-2(i-1)\right]x_i} \ldots e^{\Gamma(n-2)x_{n-1}}.
\end{equation}
There is still an exponential factor with an absolute value in its exponent. This is
\begin{equation}
e^{-\frac{\Gamma}{2}\sum_i^{n-1}|x_i-x|} = \prod_{i=1}^{n-1} \left[\theta(x_i - x)e^{-\frac{\Gamma}{2}(x_i - x)} + \theta(x - x_i) e^{\frac{\Gamma}{2}(x_i - x)} \right],
\end{equation}
where $\theta(x)$ is Heaviside's step function. In general, this contains $2^{n-1}$ terms when fully expanded, but most of these terms integrate to zero because the integral is over $x_1 > x_2 >\ldots >x_{n-1}$. In fact, there are only $n$ terms that contribute,
\begin{equation}
\begin{split}
e^{-\frac{\Gamma}{2}\sum_i^{n-1}|x_i-x|} &\rightarrow \theta(x>x_1>x_2>\ldots >x_{n-1}) e^{-\frac{\Gamma}{2}(x-x1)}e^{-\frac{\Gamma}{2}(x-x2)} \ldots e^{-\frac{\Gamma}{2}(x-x1)}\\
& + \theta(x_1>x>x_2>\ldots >x_{n-1}) e^{\frac{\Gamma}{2}(x-x1)}e^{-\frac{\Gamma}{2}(x-x2)} \ldots e^{-\frac{\Gamma}{2}(x-x1)} 
\\
& + \theta(x_1>x_2>x>\ldots >x_{n-1}) e^{\frac{\Gamma}{2}(x-x1)}e^{\frac{\Gamma}{2}(x-x2)} \ldots e^{-\frac{\Gamma}{2}(x-x1)}
\\
&+ \ldots +
\\
& + \theta(x_1>x_2>\ldots >x_{n-1}>x) e^{\frac{\Gamma}{2}(x-x1)}e^{\frac{\Gamma}{2}(x-x2)} \ldots e^{\frac{\Gamma}{2}(x-x1)},
\end{split}
\end{equation}
where the notation $\theta(x>x_1>x_2>\ldots >x_{n-1}) =  \theta(x-x_1)\theta(x_1-x_2)\theta(x_2-x_3) \ldots \theta(x_{n-2} - x_{n-1})$.

This then gives $n$ many-body integrals each over $n-1$ variables,
\begin{equation}
\begin{split}
{}_{n-1}\langle B^E | \hat{a}(x) | B^E \rangle_n &= \frac{\sqrt{\Gamma}}{2\pi}\Gamma^{n-2} (n-2)! (n-1)! e^{i E x/n} \int_> dx_1 dx_2 \ldots dx_{n-1} e^{i K_n (x_1+x_2+\ldots + x_n)/n}
\\
& \times e^{-\Gamma(n-2)x_1} e^{-\Gamma(n-4)x_2}  \ldots e^{-\Gamma\left[n-2 n-2-2(k-1)\right] x_k} \ldots e^{\Gamma(n-2)x_{n-1}}
\\
&\left\{ \theta(x-x_1) e^{-\frac{\Gamma}{2}(n-1)x} e^{\frac{\Gamma}{2}\sum_{i=1}^{n-1} x_i} + \theta(x_1-x)\theta(x-x_2) e^{-\frac{\Gamma}{3}(n-1)x} e^{-\frac{\Gamma}{2}x_1} e^{\frac{\Gamma}{2}\sum_{i=2}^{n-1} x_i} \right.
\\
& \hspace{17mm} + \ldots + \theta(x_{j-1}-x)\theta(x-x_{j}) e^{-\frac{\Gamma}{2} \left[ n-1-2(j-1)\right] x} e^{-\frac{\Gamma}{2} \sum_{i=1}^{j-1} x_i} e^{\frac{\Gamma}{2} \sum_{i=j}^{n-1} x_i}
\\
& \hspace{58mm} \left.+ \ldots + \theta(x_{n-1} - x) e^{\frac{\Gamma}{2}(n-1)x}e^{-\frac{\Gamma}{2}\sum_{i=1}^{n-1}x_i} \right\}.
\end{split}
\end{equation}
The $j$-th integral is
\begin{equation}
\begin{split}
I_j(x)&= e^{-\frac{\Gamma}{2}\left[ n-1 -2 (j-1) \right]} \int_> dx_1 dx_2 \ldots dx_{n-1} e^{i K_n (x_1+x_2+\ldots + x_n)/n} e^{-\Gamma(n-2)x_1} e^{-\Gamma(n-4)x_2} e^{-\Gamma(n-6)x_3} 
\\
& \ldots e^{-\Gamma\left[n-2-2(i-1)\right]x_i} \ldots e^{\Gamma(n-2)x_{n-1}} \theta(x_{j-1} - x) \theta(x - x_j) e^{-\frac{\Gamma}{2}(x_1+x_2+\ldots+x_{j-1})} e^{-\frac{\Gamma}{2}(x_h+x_{j+1}+\ldots+x_{n-1})}
\\
&= e^{-\frac{\Gamma}{2}\left[ n-1 -2 (j-1) \right]} \int_x^\infty dx_1 e^{-\Gamma (n-2) x_1} e^{i K_n x_1/n}e^{-\frac{\Gamma}{2} x_1} \int_x^{x_1} dx_2 e^{-\Gamma (n-4) x_2} e^{i K_n x_2/n}e^{-\frac{\Gamma}{2}x_2}\\
& \times \ldots \times \int_x^{x_{j-2}}dx_{j-1} e^{-\Gamma[n-2-2(j-2)]x_{j-1}} e^{i K_n x_{j-1}/n} 
\\
&\int_{-\infty}^{x} dx_{j} e^{-\Gamma[n-2-2(j-1)]x_{j}} e^{i K_n x_{j}/n} \int_{-\infty}^{x_j} dx_{j+1} e^{-\Gamma[n-2-2j]x_{j+1}} e^{i K_n x_{j+1}/n}
\\
&\times \ldots \times  \int_{-\infty}^{x_{n-2}} dx_{n-1} e^{\Gamma(n-2)x_{n-1}} e^{i K_n x_{n-1}/n} e^{\frac{\Gamma}{2} x_{n-1}}.
\end{split}
\end{equation}

The $n-1$ integrals thus split into two sets of integrals: the first $j-1$ integrals and the second $n-j$. The first $j-1$ integrals have a form giving
\begin{equation}
\tilde{K}_{j-1}(x) = \int_{x}^\infty dx_1 e^{-\alpha_1 x_1} \int_{x}^{x_1} dx_2 e^{-\alpha_2 x_2}  \times \ldots \times \int_{x}^{x_{j-2}} dx_{j-1} e^{-\alpha_{j-1} x_{j-1}} = \frac{\exp{\left[-\sum_{i=1}^{j-1} \alpha_i x \right]}}{\prod_{i=1}^{j-1} \left(\sum_{k=1}^i \alpha_k \right)},
\end{equation}
where $\alpha_k = \Gamma\left[n-2-2(j-1) +1/2\right] - i K_n/n$ and the sums over $\alpha_k$ can be easily evaluated. The latter $n-j$ integrals must be evaluated in reverse order. They give
\begin{equation}
\tilde{I}_{n-j}(x) = \prod_{m=1}^{n-j} \frac{1}{\Gamma m \left(n-m-1/2 \right) + i m K_n/n } \, e^{i K_n (n-j) x/n} e^{\Gamma (n-j)(j-1/2)x}.
\end{equation}
Combining the two integrals we have
\begin{equation}
\begin{split}
I_j (x) &= e^{-\frac{\Gamma}{2}\left[n-1-2(j-1) \right]x}\tilde{I}_{n-j} (x) \tilde{K}_{j-1}(x)
\\
&= e^{ i K_n'(n-1)x} \prod_{m=1}^{n-j} \frac{1}{\Gamma m \left(n-m-1/2 \right) + i m K_n' } \prod_{k=1}^{j-1} \frac{1}{\Gamma k \left(n-k-1/2 \right) - i k K_n' }
\\
&\equiv e^{ i K_n'(n-1)x} I'_j (K_n'),
\end{split}
\end{equation}
where we have defined $K_n'=K_n/n$. We note that in $I_j(x)$ the exponents containing $\Gamma$ all cancel each other.

Putting everything together the matrix element can be written
\begin{equation}
{}_{n-1}\langle B^E | \hat{a}(x) | B^E \rangle_n = \frac{\sqrt{\Gamma}}{2\pi} \Gamma^{n-2}(n-2)! (n-1)! e^{i (E - E')x} \sum_{j=1}^n I'_j (K_n').
\end{equation}
Using the properties of the complex Gamma function $\operatorname{\Gamma}{(z)}$ one can express
\begin{equation}
\sum_{j=1}^n I'_j (K_n') = \frac{\pi \, \Gamma^{1-n} \operatorname{sech}{\left[\frac{K_n' \pi}{\Gamma} \right]}}{\operatorname{\Gamma}{\left(n-\frac 1 2 - i K_n'/\Gamma\right)} \operatorname{\Gamma}{\left(n-\frac 1 2 + i K_n'/\Gamma\right)}}
\end{equation}
The entire expression becomes
\begin{equation}
\label{eq:exactBoundField}
\langle \hat{a}(x) \rangle = \frac{1}{2 \sqrt{\Gamma}} \sum_n (n-2)! (n-1)! \int dE \, dE'  c^*_{n-1}(E') c_n(E)     e^{i (E - E')x} \frac{ \operatorname{sech}{\left[\frac{K_n' \pi}{\Gamma} \right]}}{\operatorname{\Gamma}{\left(n-\frac 1 2 - i K_n'/\Gamma\right)} \operatorname{\Gamma}{\left(n-\frac 1 2 + i K_n'/\Gamma\right)}},
\end{equation}
which is thus far an exact result without approximation.

\subsection{Recovering the result of self-induced transparency} \label{sec:recoveringSIT}

We now consider Gaussian wavepackets of bound states. Here $c_n(E) = {}_n\langle B^E | \textrm{in} \rangle$. We do not need to know the full form of $c_n(E)$, but it is separable and can be written as
\begin{equation}
c_n(E) = e^{-(E - nk_0)^2 \sigma^2/(2n)} c_n,
\end{equation}
where $c_n$ is not known but is constrained by the normalization of the state $\langle \psi | \psi \rangle = 1$. Using this expression in (\ref{eq:exactBoundField}) gives
\begin{equation}
\begin{split}
\label{eq:StartOfSIT}
\langle \hat{a}(x) \rangle =& \frac{1}{2 \sqrt{\Gamma}} \sum_n (n-2)! (n-1)! c_n c_{n-1}^* \int dE \, dE'  e^{-(E' - (n-1)k_0)^2 \sigma^2/[2(n-1)]}  e^{-(E - nk_0)^2 \sigma^2/(2n)}      e^{i (E - E')x} \\
&\times \frac{ \operatorname{sech}{\left[\frac{K_n' \pi}{\Gamma} \right]}}{\operatorname{\Gamma}{\left(n-\frac 1 2 - i K_n'/\Gamma\right)} \operatorname{\Gamma}{\left(n-\frac 1 2 + i K_n'/\Gamma\right)}}.
\end{split}
\end{equation}
Again, as in the previous section, the integral can be recast over $K_n'=E/n - E'/(n-1)$ and $\bar{K}_n = E/n + E'/(n-1)$ with Jacobian $n(n-1)/2$ giving
\begin{equation}
\begin{split}
\langle \hat{a}(x) \rangle =& \frac{1}{4 \sqrt{\Gamma}} \sum_n n! (n-1)! c_n c_{n-1}^* \int dK_n' \, d\bar{K}_n  e^{-(\bar{K}_n-2k_0)^2 n \sigma^2/4} e^{-K_n^{'2} n \sigma^2/4} e^{i K_n' n x} \\
&\times \frac{ \operatorname{sech}{\left[\frac{K_n' \pi}{\Gamma} \right]}}{\operatorname{\Gamma}{\left(n-\frac 1 2 - i K_n'/\Gamma\right)} \operatorname{\Gamma}{\left(n-\frac 1 2 + i K_n'/\Gamma\right)}},
\end{split}
\end{equation}
where we have used the approximation $n/(n-1)\sim 1$ for large $n$ to simplify the Gaussian functions. Integrating over $\bar{K}_n$ gives
\begin{equation}
\begin{split}
\langle \hat{a}(x) \rangle =& \frac{1}{4 \sqrt{\Gamma}} \sum_n c_n c_{n-1}^* \frac{2 \sqrt{\pi}}{\sqrt{n \sigma^2}}\int dK_n' \,  e^{-K_n'^2 n \sigma^2/4} e^{i K_n' n x} \\
&\times \frac{ n! (n-1)! \operatorname{sech}{\left[\frac{K_n' \pi}{\Gamma} \right]}}{\operatorname{\Gamma}{\left(n-\frac 1 2 - i K_n'/\Gamma\right)} \operatorname{\Gamma}{\left(n-\frac 1 2 + i K_n'/\Gamma\right)}},
\end{split}
\end{equation}
Again, we make the approximation in the limit of large $n$,
\begin{equation}
\frac{n! (n-1)!}{\operatorname{\Gamma}{\left(n-\frac 1 2 - i K_n'/\Gamma\right)} \operatorname{\Gamma}{\left(n-\frac 1 2 + i K_n'/\Gamma\right)}} = n^2 + O(n),
\end{equation}
which leads to 
\begin{equation}
\langle \hat{a}(x) \rangle = \frac{1}{2 \sqrt{\Gamma}} \sum_n c_n c_{n-1}^* \frac{n^2 \sqrt{\pi}}{\sqrt{n \sigma^2}}\int dK_n' \,  e^{-K_n'^2 n \sigma^2/4} e^{i K_n' n x} \operatorname{sech}{\left[\frac{K_n' \pi}{\Gamma} \right]}.
\end{equation}
If the Gaussian function is slowly varying with respect to the sech function, one can expand it about $K_n'=0$  giving $1 + O(K_n')$. Taking this limit and evaluating the Fourier integral gives  
\begin{equation}
\label{eq:SITelectricFieldSUMS}
\langle \hat{a}(x) \rangle = \frac{\sqrt{\Gamma}}{2} \sum_n c_n c_{n-1}^* \frac{n^2 \sqrt{\pi}}{\sqrt{n \sigma^2}} \operatorname{sech}{\left[\frac{n \Gamma x}{2} \right]}.
\end{equation}
Defining the integration Rabi frequency as $\Omega = 2 \sqrt{\Gamma} \int dx \langle \hat{a}(x) \rangle$, one obtains
\begin{equation}
\Omega = 2 \pi \sum_n c_n c_{n-1}^* \frac{\sqrt{n \pi}}{\sigma}.
\end{equation}
The normalization of the wavefunction implies $1 = \sum_n \int dE |c_n(E)|^2 = \sum_n |c_n|^2 \frac{\sqrt{n \pi}}{\sigma}$ and therefore the integrated Rabi frequency in the large $n$ limit approaches
\begin{equation}
\Omega = 2 \pi.
\end{equation}
From (\ref{eq:SITelectricFieldSUMS}), in the limit of a large average photon number and a narrow photon distribution,  the electric field can also be approximated as
\begin{equation}
\label{eq:electricFieldSIT}
\langle \hat{a}(x) \rangle = \frac{\bar{n} \sqrt{\Gamma}}{2}\operatorname{sech}{\left(\frac{\bar{n} \Gamma x}{2} \right)},
\end{equation}
where $\bar{n}$ is the average photon number.

\section{Normally-ordered correlation function of bound states and mean-field theory}

Here we compute the $m$-th-order normally ordered correlation function of the bound states ($n \geq m$). Following the previous section one obtains an integral
\begin{equation}
\begin{split}
{}_n\langle B^{E'} | \left[\hat{a}^\dagger(x) \right]^m  \left[\hat{a}(x) \right]^m | B^E \rangle_{n} &= \frac{\Gamma^{n-1} (n-1)!}{2 \pi n} \frac{n!}{(n-m)!} \int dx_1 dx_2 \ldots dx_{n-m} e^{i (E - E')(x_1+x_2+\ldots x_{n-m})/n}\\
& \times e^{-\Gamma \sum_{i<j}^{n-m} |x_i - x_j|} e^{i E m x /n} \prod_{i=1}^{n-m} e^{-m \Gamma |x_i - x|}.
\end{split}
\end{equation}
As before, one must reduce the integral to the domain $x_1 > x_2 > \ldots > x_{n-m}$ and then expand the product and reduce it to its $n-m+1$ non-zero terms. Computing all the integrals using the techniques of the previous section, one obtains
\begin{equation}
\begin{split}
\label{eq:mthOrderBoundNMatrixElement}
{}_n\langle B^{E'} | \left[\hat{a}^\dagger(x) \right]^m  \left[\hat{a}(x) \right]^m | B^E \rangle_n &= \frac{\Gamma^{m-1}}{2 \pi n} \left(\frac{n!}{(n-m)!} \right) \frac{(n-1)! (n+m-1)!}{(2m-1)!} \frac{\operatorname{\Gamma}{(1-n- i \frac{K}{n\Gamma})} \operatorname{\Gamma}{(1-n+ i \frac{K}{n\Gamma})} }{\operatorname{\Gamma}{(1-m- i \frac{K}{n\Gamma})} \operatorname{\Gamma}{(1 - m + i \frac{K}{n\Gamma})}} e^{i K x},
\end{split}
\end{equation}
where now $K = E - E'$.

\subsection{Reduction to mean-field theory $(n \gg m)$}

We now show that if the particle number $n$ is much large than the order of the correlation function $m$, the outcome of the measurement probes the mean fields and the correlation function factorizes. One can start by making the approximation
\begin{equation}
\begin{split}
(n!)^2  \operatorname{\Gamma}{\left(1-n- i \frac{K}{n\Gamma} \right)} \operatorname{\Gamma}{\left(1-n+ i \frac{K}{n\Gamma}\right) }  = \frac{\pi^2 n^2}{\operatorname{sinh}^2{(\frac{\pi K}{n \Gamma})}} + O(n).
\end{split}
\end{equation}
Additionally one can use the recurrence relation of the Gamma function $\Gamma(z-n) = \Gamma(z)/\left[ (z-1)(z-2) \ldots (z-n) \right]$ to write
\begin{equation}
\begin{split}
\operatorname{\Gamma}{\left(1-m- i \frac{K}{n\Gamma} \right)} \operatorname{\Gamma}{\left(1-m+ i \frac{K}{n\Gamma}\right) } &= \operatorname{\Gamma}{\left(-\frac{i K}{n\Gamma}\right) } \operatorname{\Gamma}{\left(\frac{i K}{n\Gamma}\right) } \prod_{k=1}^{m-1} \frac{1}{k^2 + \frac{K^2}{n^2 \Gamma^2}}\\
&= \frac{\pi}{\frac{K}{n \Gamma} \operatorname{sinh}{\left(\frac{\pi K}{n \Gamma}\right)}} \prod_{k=1}^{m-1} \frac{1}{k^2 + \frac{K^2}{n^2 \Gamma^2}}. 
\end{split}
\end{equation}
Combining these, the matrix element becomes
\begin{equation}
\begin{split}
{}_n\langle B^{E'} | \left[\hat{a}^\dagger(x) \right]^m  \left[\hat{a}(x) \right]^m | B^E \rangle_n &= \frac{\Gamma^{m-1}}{2} \frac{n^{2m-1}}{(2m-1)!} \frac{K}{n \Gamma} \operatorname{cosech}{\left( \frac{\pi K}{n \Gamma} \right)} e^{i K x}\prod_{k=1}^{m-1} \left( k^2 + \frac{K^2}{n^2 \Gamma^2} \right),
\end{split}
\end{equation}
where we have used Stirling's approximation to write $\frac{(n+m-1)!}{(n-m)!} \sim n^{2m-1}$.

One can now consider an $n$-photon Fock state composed of bound states
\begin{equation}
\label{eq:psiBoundFockSIT}
|\psi \rangle_n = \left(\frac{\sigma}{\sqrt{n \pi}} \right)^{1/2} \int dE \, e^{-(E-n k_0)^2 \sigma^2/(2n)} |B_n^E \rangle.
\end{equation}
The correlation function of this state is given by 
\begin{equation}
{}_n\langle \psi | \left[\hat{a}^\dagger(x) \right]^m \left[\hat{a}(x) \right]^m |\psi \rangle_n = \frac{\sigma}{\sqrt{n \pi}} \int dE dE' \, e^{-(E'-n k_0)^2 \sigma^2/(2n)} e^{-(E-n k_0)^2 \sigma^2/(2n)} \langle B_n^{E'} | \left[\hat{a}^\dagger(x) \right]^m  \left[\hat{a}(x) \right]^m | B_n^E \rangle.
\end{equation}
Changing the integration variables to $K=E - E'$ and $\bar{K}=(E+E')/2$ and integrating out $\bar{K}$ gives
\begin{equation}
{}_n\langle \psi | \left[\hat{a}^\dagger(x) \right]^m \left[\hat{a}(x) \right]^m |\psi \rangle_n = \frac{\Gamma^{m-1}}{2} \frac{n^{2m-1}}{(2m-1)!} \int dK \frac{K}{n \Gamma} \operatorname{cosech}{\left( \frac{\pi K}{n \Gamma} \right)} e^{-\frac{K^2 \sigma^2}{4 n}} e^{i K x}\prod_{k=1}^{m-1} \left( k^2 + \frac{K^2}{n^2 \Gamma^2} \right).
\end{equation}
At this point one can make use of the tabulated Fourier Transform \cite{BatemanBookIntegralTransforms},
\begin{equation}
[\operatorname{sech}{(ax)}]^{2n} = \int dy \, e^{i y x} \frac{4^{n-1}}{ (2n-1)! a} \frac{y}{2a} \operatorname{cosech}{\left(\frac{\pi y}{2 a} \right)} \prod_{r=1}^{n-1}\left(\frac{y^2}{4a^2} + r^2\right).
\end{equation}
In the limit $\sigma \Gamma \ll 1$ the Gaussian factor approaches unity and the tabulated integral can be used to give
\begin{equation}
\begin{split}
{}_n\langle \psi | \left[\hat{a}^\dagger(x) \right]^m \left[\hat{a}(x) \right]^m |\psi \rangle_n &= \frac{\Gamma^{m-1} n^{2m-1}}{2}  \frac{n\Gamma}{2} \frac{1}{4^{m-1}} \left[\operatorname{sech}{\left( \frac{n \Gamma x}{2} \right)} \right]^{2m}\\
& = \left[ \frac{n \sqrt{\Gamma}}{2} \operatorname{sech}{\left( \frac{n \Gamma x}{2} \right)} \right]^{2m}.
\end{split}
\end{equation}
As in the previous section, one can also consider a state composed of different Fock states $n$. Such a state returns the same result but with $n$ replaced with the average photon number $\bar{n}$. Comparing this result with (\ref{eq:electricFieldSIT}) indicates that the $m$-th order correlation function can be obtained using mean-field theory when $n \gg m$. We note that the expression for the correlation function has a different form when $n \sim m$. This is particularly clear from (\ref{eq:mthOrderBoundNMatrixElement}) when $n=m$ which causes the Gamma functions to evaluate to unity leaving only the plane-wave factor varying with $K$.

\section{Propagation in the large $n$ limit}

In this section we show how the observables change under propagation when $n \gg 1$. We start with the electric field $\langle \hat{a}(x) \rangle$ which is given in (\ref{eq:exactBoundField}). For an incident coherent field in the limit of $n \gg 1$, we expect that the field amplitudes of the bound states approach those of a coherent state and therefore 
\begin{equation}
c_n(E) \sim e^{-|\alpha|^2/2}\frac{\alpha^n}{\sqrt{n!}} \left(\frac{\sigma}{\sqrt{n \pi}} \right)^{1/2} e^{-(E - n k_0)^2 \sigma^2/(2n)}
\end{equation}
Since the bound states are eigenstates, scattering off $N$ emitters simply maps $c_n(E) \rightarrow t_{E,n}^N c_n(E)$. As in the main text, one can write $t_{E,n}^N = e^{i N \phi_n(E)}$ and Taylor expand $\phi_n(E)$. Since we have found the pulse distortion terms decrease rapidly as $n$ increases we simply consider the first order term in the expansion of $\phi_n(E)$. Following the calculations starting from  (\ref{eq:StartOfSIT}) one obtains for zero detuning ($k_0=0$)
\begin{equation}
\langle \hat{a}(x) \rangle  = e^{-|\alpha|^2} \sum_n \frac{\alpha^{2n}}{n!} \frac{n\sqrt{\Gamma}}{2} \operatorname{sech}{\left[ \frac{n \Gamma}{2} \left(x + \frac{4 N}{n^2 \Gamma} \right) \right]},
\end{equation}
which gives (\ref{eq:SITbeyondMF}) from the main text. A similar process can be followed to obtain the normally ordered correlation functions.

\section{Effective spin model and MPS ansatz}\label{sec:ME}
In this appendix we present an alternative route to the S-matrix scattering formalism used in the main text to describe the light propagation through an array of  chirally coupled emitters. The full system  dynamics   can be indeed described by the combination of a driven-dissipative  master equation for the emitters and a generalized input-output relation   that allowed to reconstruct the electric field~\cite{Caneva2015NJP, Manzoni2017NCOM}. The dynamics of the emitter can than be efficiently solved by using a matrix product states (MPS) ansatz~\cite{Manzoni2017NCOM,Bienias2018arXiv,Verstraete2008ADPHYS,Schollwock2011Annals} which allow us to push the simulation to larger atomic arrays and higher input power.

\subsection{Effective spin model}
\begin{figure}[!t]
\includegraphics[width=0.6\columnwidth]{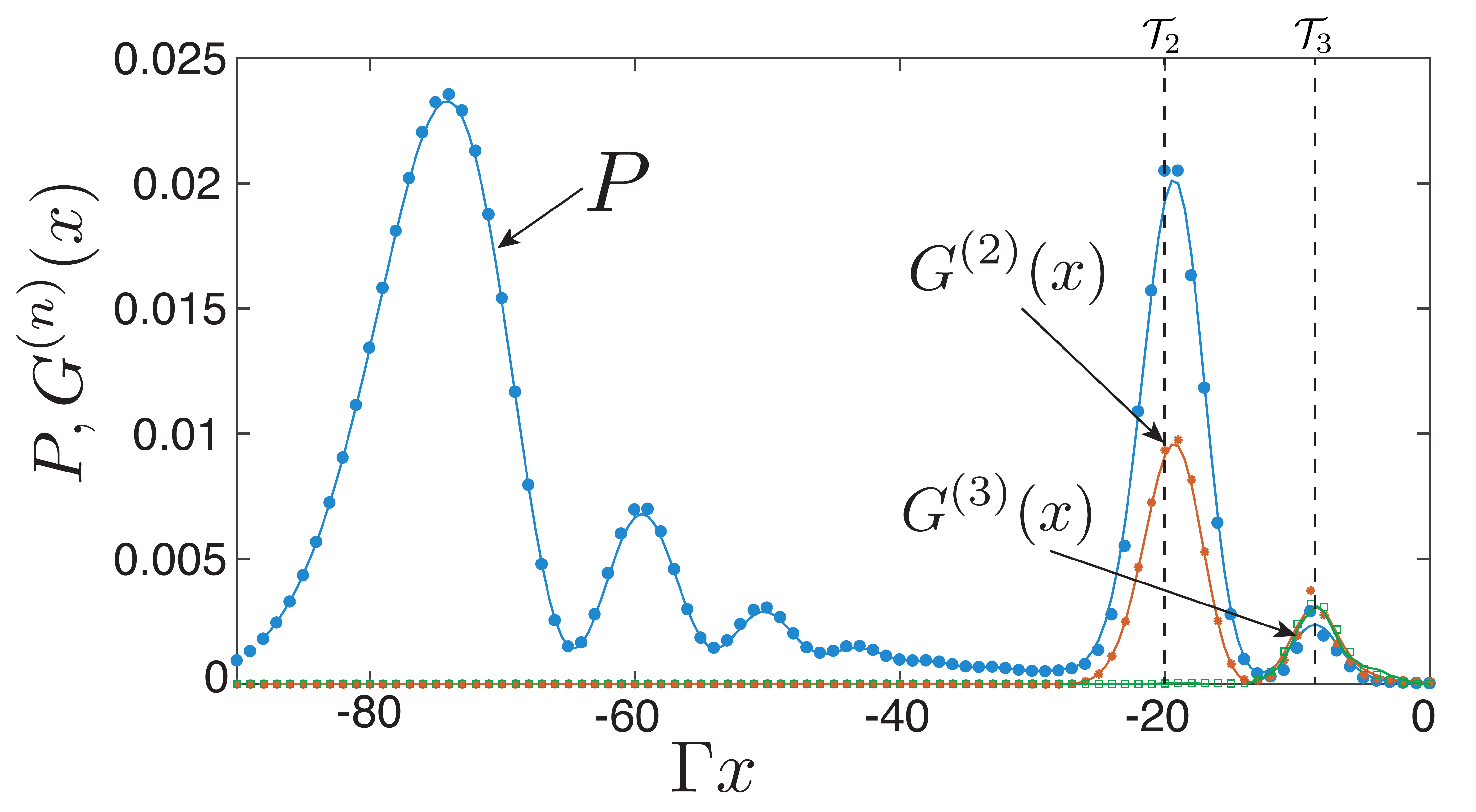}
\caption{\label{fig.supp1} Comparison between the results obtained by directly solving the master equation (continuous line) and by using  the quantum trajectories algorithm (markers points) for the same parameters used in Fig.2(d) of the main text. }
\end{figure}
Let us consider the most general case (extension of the cascade scenario described by Eq.(1) of the main text) of $N$ TLA 
 with equal frequency $\omega_0=ck_0$ coupled asymmetrically  to a waveguide with decay rates $\Gamma_R$ and $\Gamma_L$ associated respectively to the emission of right and left propagating photons. The addition of the two decay rates  gives the total emission  rate of the TLA into the waveguide $\Gamma=\Gamma_L+\Gamma_R$. In a realistic implementation the emitters can also radiate into  other external channels, the decay rate associated to this emission is indicated with $\Gamma_0$.
The right propagating input pulse is treated as a classical coherent field, $E_{in}(t,x)=\mathcal{E}_{in}(t)e^{ik_{in} x}$ and for the rest of the discussion we will assume it to be on resonance with the atomic transition, i.e. $k_{in}=k_0$.
The  coupling of the emitters to the input field is given  by the Hamiltonian $H_{\rm drive}=\sum_j\sqrt{\Gamma_R}\left[E_{in}(t,x_j)\hat\sigma_{j}^++ H.c.\right]$ where $\hat\sigma_{j}^+=|g\rangle\langle e|_j$ is the $j$-th  emitter annihilation operator.

Under Born-Markov approximation the emitters dynamics is known to be described by a chiral master equation(ME) of the general form~\cite{Pichler2015PRA}:
\begin{equation}\label{eq:MasterEq_supp}
\dot \rho= -i \left( H_{\rm eff}\rho-\rho H^{\dagger}_{\rm eff}\right)+\mathcal{J}[\rho],
\end{equation}
where 
\begin{equation}\label{eq:Heff_supp}
\begin{split}
H_{\rm eff}=-i\sum_j\frac{\left(\Gamma+\Gamma_0\right)}{2}\hat\sigma_{j}^+\hat\sigma_{j}^-+H_{\rm drive}
-i\sum_{l>j}\left(\Gamma_Le^{ik_0|x_l-x_j|}\hat\sigma_{j}^+\hat\sigma_{l}^-+\Gamma_Re^{ik_0|x_l-x_j|}\hat\sigma_{l}^+\hat\sigma_{j}^-\right)
\end{split}
\end{equation}
is the effective  Hamiltonian which provides the non-Hermitian collective evolution of the emitters.   The recycling term 
 \begin{equation}
\begin{split}
\mathcal{J}[\rho]=\left(\Gamma+\Gamma_0\right)\sum_j\hat\sigma_{j}^-\rho \hat\sigma_{j}^+
+\sum_{l>j}\left[\left(\Gamma_Re^{ik_0|x_l-x_j|}+\Gamma_Le^{-ik_0|x_l-x_j|}\right)\hat\sigma_{j}^-\rho\hat\sigma_{l}^++\rm H.c.\right]
\end{split}
\end{equation}
assures the conservation of the density operator trace
and it arises by the quantum  jumps that the emitters experience after the   emission of a photon  into the waveguide or into free space. 
For $\Gamma_L=\Gamma_R$ Eq.~\eqref{eq:MasterEq_supp}  coincides with  the standard waveguide QED master equation~\cite{Chang2012NJP,Gonzalez-Tudela2011PRL} where both coherent and dissipative interactions between the emitters occur depending on the atoms position.
In the limit of a perfect chiral waveguide considered in the main text, i.e.  $\Gamma_L=\Gamma_0=0$ and $\Gamma=\Gamma_R$, the master equation Eq.~\eqref{eq:MasterEq_supp} reduces to the well known cascade master equation~\cite{Stannigel2012NJP,Pichler2015PRA} characterized by an unidirectional purely dissipative interaction between the emitters. 
In this limit only the spatial order of the atoms and not the specific
positions matters and  the excitation is transferred only to the right without any back action. \\
Once solved the emitters dynamics, the transmitted and reflected (for the semi-chiral case) electric field is obtained by  the following generalized input-output relations~\cite{Caneva2015NJP, Manzoni2017NCOM}:
\begin{equation}\label{in_outR_supp}
E_R(t)=\mathcal{E}_{\rm in}(t)+i\sum_j\sqrt{\Gamma_R}e^{-ik_0x_j}\hat\sigma_{j}^-(t),
\end{equation}  
\begin{equation}\label{in_outL_supp}
E_L(t)=i\sum_j\sqrt{\Gamma_L}e^{ik_0x_j}\hat\sigma_{j}^-(t).
\end{equation}  
  The combination of Eq.~\eqref{eq:MasterEq_supp}, \eqref{in_outR_supp}, and \eqref{in_outL_supp} allows to correctly describes the photon propagation trough the chiral medium. 

\subsection{MPS ansatz}
In order to observe the main results obtained in the main text, e.g. photon number dependent bound states separation, it is necessary to consider substantial input pulse amplitudes  and large atomic arrays, a scenario that is  challenging to simulate with standard numerical techniques. To treat this many body problem we use a recently developed algorithm which involve an MPS representation to efficiently describe the effective spin dynamics~\cite{Manzoni2017NCOM}.  
The system evolution can be solved with two different approaches, either by directly solving the ME~\eqref{eq:MasterEq_supp}~\cite{Bienias2018arXiv}(method used for Fig.~3 of the main text) or by using a quantum trajectories algorithm where the state of the system evolves under the effective Hamiltonian~\eqref{eq:Heff_supp} and it stochastically experiences the occurring of quantum jumps~\cite{Manzoni2017NCOM} (method used for Fig.~2(f) of the main text). In both case an MPS representation is applied either to the quantum state or to the 
the density matrix once mapped to a vector. The MPS ansatz consists in 
reshaping the generic quantum state $|\phi/\rho\rangle=\sum_{i_1,..i_N}\psi_{i_1,i_2,..i_N}|i_1,i_2,..i_N\rangle$ into a matrix product state of the form:
\begin{equation}
|\phi/\rho\rangle=\sum_{i_1,..i_N}A_{i_1}A_{i_2}...A_{i_N}|i_1,i_2,..i_N\rangle
\end{equation}
where, for each specific set of physical indices $\{\bar i_1,\bar i_2,..\bar i_N\}$, the product of the $A_{\bar i_j}$ matrices gives back the state coefficient $\psi_{\bar i_1,\bar i_2,..\bar i_N}$. Each matrix $A_{\bar i_j}$ has dimension $D_{j-1}\times D_{j}$ known as bond dimension and finite-edge boundary conditions are assumed by imposing $D_{1}=1$ and $D_{N}=1$. The bond dimension reflects the entanglement entropy. For arbitrary states the bond dimensions grow exponentially with the system size. The advantage of the MPS ansatz it is that, in many physical scenarios, as the one considered here, the entanglement can grow slowly with the system size allowing an efficient description of the state in terms of a smaller bond dimension~\cite{Schollwock2011Annals}. 

In order to compute the evolution of the system it is possible, in both methods, to derive a  matrix product  operator (MPO) representation for either the Liouvillian or the effective Hamiltonian and jump operators. The derivation of this representation for our semi-chiral case follows straightforwardly from the bi-directional case presented in~\cite{Manzoni2017NCOM,Bienias2018arXiv}. The evolution of the system is then computed by applying a linear expansion of the master equation (or of the time evolution operator) $1+dt\mathcal{L}(H_{\rm eff})$ or a Runge-Kutta  method. In addition a MPO representation can be derived also for the observables and it allows to evaluate the expectation values.  After each application of an  MPO  to an MPS  the MPS bond dimension increases. The bond dimension is than truncated after each step in order to keep an efficient description of the state. We indicate as $D_{max}$ the maximum bond dimension used to represent the system during an entire time evolution. In all the plots considered in this work we used $D_{max}=150$ for the ME simulations and $D_{max}=40$ for the quantum trajectories algorithm.

 \section{Effect of other decay channels}\label{sec:decay}

\begin{figure}[!t]
\includegraphics[width=0.8\columnwidth]{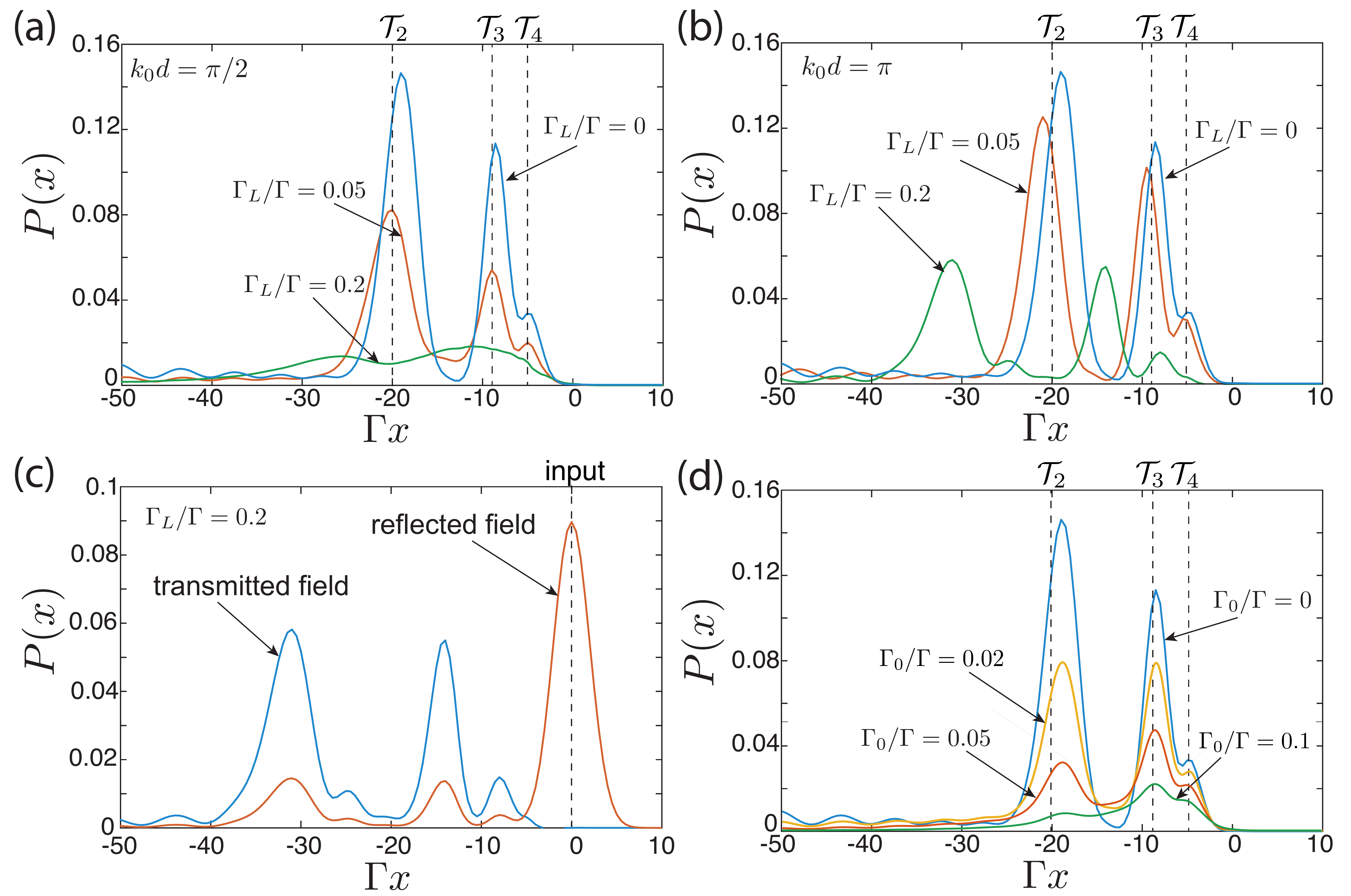}
\caption{\label{fig_supp_2} (a)-(b) Transmitted power as function of the position for different values of  the decay along the left direction. In (a) we fixed the atomic distance to $kd=\pi/2$ while in (b) to $kd=\pi$. (c) Transmitted and reflected power for $\Gamma_L/\Gamma=0.2$ and $kd=\pi$. (d) Transmitted power for different values of an external decay rate $\Gamma_0$. In panels (a),(b) and (d) the dashed lines show the expected delay of the BS for the $\Gamma_L=\Gamma_0=0$ case while in panel (c) it shows the input field position. In all panels $N=20$, $\bar n=1.8$, $\Gamma\sigma=2\sqrt{2}$ and $D_{\rm max}=170$.}
\end{figure}

In the main text we considered the ideal scenario of an array of atoms perfectly chirally coupled to a waveguide. In real implementations this is not always the case and the atoms can either emit into the other direction or into  free space. 
Let us first assume that the atoms can emit also into the direction opposite to the pulse propagation,  i.e. $\Gamma_L\ne 0$, while the emission to other external channels is suppressed $\Gamma_0=0$. In this case the master equation~\eqref{eq:MasterEq_supp} becomes dependent on the atomic positions. To better understand this point we consider the two paradigmatic cases of atoms equally spaced either by a distance $k_0d=k_0|x_i-x_j|= (n+1)\pi/2$ or  $k_0d=k_0|x_i-x_j|= n\pi$ with $n=1,2,..$. We focus on these two configurations because  in the bi-directional limit they  are the ones providing purely coherent or dissipative interactions respectively. In Fig.\ref{fig_supp_2}(a)-(b) we plot the transmitted intensity for the two cases and we compare it to the one obtained in the perfectly chiral scenario. We observe that the emission into the opposite channel affects the intensity in a totally different manner depending on the atomic distance. In the first configuration, Fig.\ref{fig_supp_2}(a), it leads to an overall  spreading of the bound states peaks for increasing values of $\Gamma_L$. In the second, Fig.\ref{fig_supp_2}(b), the bound state separation not only  seems to be robust to the deviation from a perfect chiral emission, but it even leads to a better resolution of the bound state peaks at the price of a weaker pulse in transmission. This effect is caused by an additional time-delay of the BS  peaks, induced by  partial reflection, which affects mostly the few photon bound states making the bound states separation  even more evident.  The many-photons BS, on the contrary, not only experience less delay than their few-photons counterpart, but they are also more robust against distortion tending to keep  their solitonic behaviour.
These features  are more evident in Fig.~\ref{fig_supp_2}(c) where we  plot the reflected field intensity. Here the majority of the reflection occurs at the interface and it causes the damping in transmission. On the other end,  when the pulse enter in the medium, we observe a major reflection for the BS with few photons. This phenomena make sense  in light of the connection to the SIT limit at large photon pulses. In this limit, the chirality  is not  an essential  requirement to observe the formation of solitons. The role of chirality is however crucial to bring the solitonic behaviour to the level of few photons.

In addition this observation suggests that the $k_0d=k_0|x_i-x_j|= n\pi$ spacing is the optimal configuration to observe the effect in an experiment. 
On the other hand,  in many realistic implementation, it is difficult to have control on the position of the emitters. For this reason it makes sense consider the effect of a semi-chiral emission by making an average on random positions. This scenario is similar to consider a decay of the atoms in an external channel with rate $\Gamma_0$. The effect of this external decay is plotted in Fig.~\ref{fig_supp_2}(d) and shows that in general the effect is visible for values  $\Gamma_0\le 0.2$. Again we expect the overall effect to be robust for many-photon BS.

\end{document}